\documentclass{aastex63}
\usepackage{xcolor}
\usepackage{float}
\makeatletter
\let\newfloat\newfloat@ltx
\makeatother
\usepackage{graphicx}
\usepackage{amsmath}
\usepackage{enumerate}
\usepackage{natbib}

\usepackage{appendix}
\usepackage{algorithm}
\usepackage[noend]{algpseudocode}
\graphicspath{{./}{figures/}}

\submitjournal{ApJ}

\shorttitle{Exploration about the origin of star clusters}
\shortauthors{Chattopadhyay et al.}

\begin{document}

\title{ EXPLORATION ABOUT THE ORIGIN OF GALACTIC AND EXTRAGALACTIC STAR CLUSTERS THROUGH SIMULATED H-R DIAGRAMS}

\author{Tanuka Chattopadhyay}
\email{tchatappmath@caluniv.ac.in}
\author{Sreerup Mondal}
\author{Suman Paul}
\affiliation{Department of Applied Mathematics \\ University of Calcutta \\
 92, A.P.C Road Kolkata 700009, India}
\author{Subhadip Maji}
\affiliation{Indian Statistical Institute, Kolkata \\
203, Barrackpore Trunk Rd, Dunlop \\
Kolkata 700108, India}
\author{Asis Kumar Chattopadhyay}
\affiliation{Department of Statistics\\ University of Calcutta \\
35, Ballygunge Circular Road \\
Kolkata 700019, India}

\correspondingauthor{SUMAN PAUL}
\email{spappmath$\_rs$@caluniv.ac.in}

\begin{abstract}

The present work explores the origin of the formation of star clusters in our Galaxy and in Small Magellanic Cloud (SMC) through simulated H$-$R diagrams and compare those with  observed star clusters. The simulation study produces synthetic H$-$R diagrams by Markov Chain Monte Carlo (MCMC) technique using star formation history (SFH), luminosity function (LF), abundance of heavy metal (Z) and a big library of isochrones as basic inputs and compares them with observed H$-$R diagrams of various star clusters. The distance based comparison between those two diagrams is carried out through two dimensional matching of points in Color$-$Magnitude Diagram (CMD) after optimal choice of bin size and appropriate distance function. It is found that a poor medium of heavy elements (Z $=$ 0.0004), Gaia LF along with mixture of multiple Gaussian distributions of SFH may be the origin of formation of globular clusters (GCs). On the contrary, enriched medium (Z $=$ 0.019) is favoured with Gaia LF along with double power law (i.e. unimodal) SFH. For SMC clusters, the choice of exponential LF and exponential SFH is a proper combination for poor medium whereas Gaia LF with Beta type SFH is preferred in an enriched medium for the formation of star clusters.

\end{abstract}


\section{Introduction} \label{sec:intro}

Hertzsprung$-$Russel (H$-$R) diagram is a bivariate plot of absolute magnitude versus temperature or colour for a large number of stars in resolved stellar populations or galaxies. This diagram provides a snapshot of the evolutionary status of the stars, bright enough to be detected. Various parameters like star formation histories (SFH), luminosity functions (LF), chemical abundance or metallicity (Z) etc, interplay, in a significant way which in turn determines the shape of H$-$R diagram in composite stellar populations. The stellar populations are much more complex systems containing numerous different star formation epochs superimposed upon a single H$-$R diagram. Thus the H$-$R diagrams or color-magnitude diagrams (CMD) require more sophisticated techniques for accurate interpretation.\\

The most common technique to interpret a CMD is to produce a series of isochrones (stars which are at the same time and metallicity of their evolutionary status). These will match as many characteristics of the diagram as possible and thus, are either older or younger in age than the majority of the stars (Miller et  al. \citeyear{Miller2001}, J{\o}rgensen and Lindegren \citeyear{Jorgensen2005}, Monteiro et al. \citeyear{Monteiro2010}). Also this verification method is appropriate for CMD of star clusters, created at a single point of time (Sandage \citeyear{Sandage1953}, Sandage \citeyear{Sandage1958}, Stetson \citeyear{Stetson1993}, Kalirai and Tosi \citeyear{Kalirai2004}). But recent studies show that resolved stellar populations have not been originated at a single epoch but at multiple epochs (Mackey et al. \citeyear{Mackey2008}, Katz and Ricotti \citeyear{Katz2013}, Bastian and Lardo \citeyear{Bastian2018}). Thus a possible way to properly interpret a complex CMD is through statistical (Monte Carlo) simulation where a composite stellar population can be generated from evolutionary tracks using LF, SFH and metallicities and matched with the observed ones to properly interpret their origin. The above discussion is the motivation behind the present work. Similar idea was first applied to galaxies (Ferraro and Fusi Pecci \citeyear{Ferraro1989})  and  elaborate models were developed  (Tosi et al. \citeyear{Tosi1991}, Bertelli et al. \citeyear{Bertelli1992}, Greggio et al. \citeyear{Greggio1993}, Hernandez et al. \citeyear{Hernandez1999}). The unveiling of unknown SFH incorporating measurement errors was studied by Tolstoy and Saha (\citeyear{Tolstoy1996}). Parametric models for galaxy SFHs were developed by Carnall et al. (\citeyear{Carnall2019}) and the role of various SFH were studied in photometry of these objects. Attempts have been made to make comparisons between synthetic CMDs and observed CMDs by Arp (\citeyear{Arp1967}), Robertson (\citeyear{Robertson1974}), Harris and Deupree (\citeyear{Harris1976}), Flannery and Johnson (\citeyear{Flannery1982}), Becker and Mathews (\citeyear{Becker1983}), Salaris et al. (\citeyear{Salaris2007}), Fiorentino et al. (\citeyear{Fiorentino2011}), Martins and Palacios (\citeyear{Martins2017}). \\ 
  
Statistically reliable methods were also used by various authors for comparing synthetic (simulated) and observed data sets before, in X-ray astronomy (Lampton et al. \citeyear{Lampton1976}, Sarazin \citeyear{Sarazin1980}, Bradt et al. \citeyear{Bradt1992}, Ramsey et al. \citeyear{Ramsey1994}, Gruber et al. \citeyear{Gruber1999}) and in quasar distribution (Peacock \citeyear{Peacock1983}, Fasano and Franceschini \citeyear{Fasano1987}). These techniques have been used in situations where a functional form of the distributions are available. On the contrary a CMD model is a complex two dimensional non linear distribution of data points and it is important to take into consideration  the facts that the two sets of data points match with respect to spatial distribution as well as the relative number of points at different positions in the diagram.\\

In the present work we have considered SFH, LF and metallicities as the basic inputs for producing synthetic H-R diagram and then matched the synthetic diagrams to the observed ones for resolved stellar populations to explore about their origin. The present work has the following improvements over the previous works: 

\begin{itemize}

    \item We have used different SFHs instead of a single SFH.
    
    \item  We have used different LFs instead of a single one and also various power law luminosities observed in 20 star forming galaxies based on Hubble Legacy Archive Photometre (Whitmore et al. \citeyear{Whitmore2014}) and Gaia luminosity function (Brown et al. \citeyear{Brown2018}).
    
    \item  We have matched the CMD diagrams of observed and synthetic ones by computing `minimum distances'  between the bivariate histograms. This takes into consideration of both the spatial as well as probability distributions.
    
    \item We have optimized the `bin' size and `distance function' through comparison with other bin sizes and distance functions existing in the literature.
    
\end{itemize}
    
In Section \ref{sec:The Model} we have developed the mathematical model. In Section \ref{sec:Various star formation histories} the various forms of the SFHs have been discussed. Section \ref{sec:Luminosity functions} describes the LFs for both observed and theoretical ones used in the work. Section \ref{sec:5} gives a short description of the matching model. Results and discussions are demonstrated in Section \ref{Results and discussions}. Section \ref{Conclusion} outlines the conclusions.

\section{The Model} \label{sec:The Model}

To produce a synthetic CMD, we first assume a SFH i.e. SFR(t) (star formation rate as a function of time) and a luminosity function (LF). CMD requires a method of obtaining the colour and luminosity (or absolute magnitude) of a star of given mass and age. For finding colour and luminosity at a given age and metallicity, various theoretical isochrones are used. If the isochrones are largely spaced over time then interpolation between isochrones will be an erroneous procedure which can introduce spurious structure in the ultimate result. To avoid such error we use latest Padova (Fagotto et al. \citeyear{Fagotto1994}, Girardi et al. \citeyear{Girardi2000}) full stellar tracks, calculated at fine variable time intervals and a careful interpolation method is used at constant evolutionary phases to construct an isochrone library. We have constructed almost 1000 points in each isochrone, so that the luminosity resolution is very small. Also we have taken two metallicity limits as Z $=$ 0.0004 and Z $=$ 0.019 i.e. the minimum and maximum values, to explore the effect of metallicity, if any, in the origin and hidden properties of stellar populations. We fit linear splines to the isochrones at $l$, evenly spaced luminosity intervals
given a tabulated function $y_{i} = y(x_{i})$, $i=1,2,3,...,l_{j}$, $j$, being the $j$th isochrone.\\ 
For $i=1,2,3,...,l_{j}$, the interpolating function joins ($l_{j}-1$) linear functions of the form
\begin{equation}
   f_{i}(x) = a_{i}y_{i} + b_{i}y_{i+1}, \quad x \in [x_{i}, x_{i+1}] 
\end{equation}
 where $a_{i}$, $b_{i}$ are constants satisfying 
 (i) $f_{i}(x_{i}) = y_{i}$,  (ii) $ f_{i}(x_{i+1}) = y_{i+1} $,  i.e.  $a_{i} = \frac{x_{i+1}-x}{x_{i+1}-x_{i}}$  and $b_{i} = 1-a_{i} = \frac{x-x_{i}}{x_{i+1}-x_{i}}$,  $i= 1,2,...,(l_{j}-1$). \\

\begin{figure}[h]
    \centering
    \includegraphics[scale=0.9]{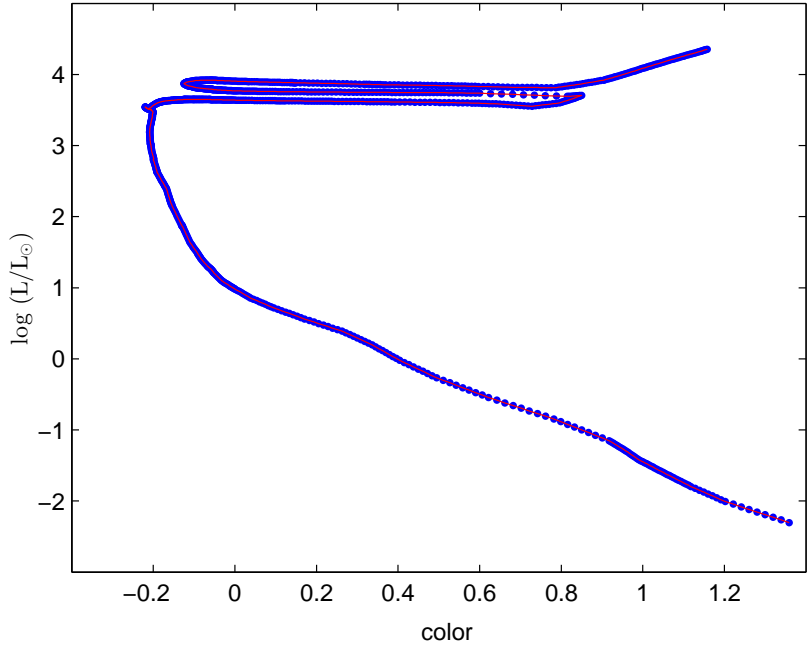}
    \caption{ Interpolated points  (1200) (in red) on a particular original isochrone (in blue) of 50 points at time $= 0.0634$ Gyr  and metallicity $Z$ = 0.0004. The luminosity ( $l$ i.e. $log(L/L_\odot)$) is along vertical axis and color (c) is along horizontal axis.} \label{fig:1}
\end{figure}

Then we start again with newly interpolated points along with the old ones and repeat the process to include more interpolating points. We repeat the same procedure for all other isochrones. This decreases the interpolation error (Cassisi et al. \citeyear{Cassisi2012}). Fig. \ref{fig:1} shows one such original isochrone along with interpolated isochrone formed using linear interpolation, iteratively used.\\

 To produce a synthetic CMD the following steps are performed:
 
 \begin{enumerate}
     \item One possible choice of probability density function (pdf) of SFR ($t$) is considered.
     
     \item One particular choice of LF is considered.
     
     \item One timepoint t is randomly generated from SFR ($t$) distribution.
     
     \item One luminosity point ($l$) is randomly generated from LF  distribution.
     
     \item Corresponding to the time ($t$) and luminosity ($l$) the corresponding colour ($c$) is extracted from the interpolated isochrone library.
     
     \item Steps (3)$-$(5) are repeated for a large number of times (5000 in our case).
     
     \item The various luminosities ($l_{i}$) and colors ($c_{i}$), $i$=1, 2,...,5000, are plotted, which is the synthetic CMD in our case.
     
     \item The synthetic CMD is compared with various observed CMD diagrams of open and globular clusters and the matched pair is chosen for which the spatial distance is minimum.
     
 \end{enumerate}

\section{Various star formation histories} \label{sec:Various star formation histories}

Following Carnall et al. (\citeyear{Carnall2019}), three SFH models are considered. They are 

\begin{itemize}

    \item Exponentially declining (Searle et al. \citeyear{Searle1973}, Pacifici et al. \citeyear{Pacifici2013}, McLure et al. \citeyear{McLure2018}, Wu et al. \citeyear{Wu2018}).
    
    \item Delayed exponentially declining (Ciesla et al. \citeyear{Ciesla2017}, Chevallard et al. \citeyear{Chevallard2019}).
    
    \item Double power law (Ciesla et al.\citeyear{Ciesla2017}, Carnall et al.
    \citeyear{Carnall2019}).

\end{itemize}

In addition we have also considered Gaussian Mixture Model (2 and 3 modes) following the episodic nature of star formation (Hernandez et al. \citeyear{Hernandez1999}, Stinson et al. \citeyear{Stinson2007}, Tremblay et al. \citeyear{Tremblay2010}, Huang et al. \citeyear{Huang2013},  Debsarma et al. \citeyear{Debsarma2016}, Cignoni et al. \citeyear{Cignoni2018}, Das et al. \citeyear{Das2020}) and Beta distribution following  unimodal star formation history in many giant galaxies.

\subsection{Exponentially declining SFR(t)} \label{sec:Exponentially declining SFR(t)}

Exponentially declining SFHs are widely used model of SFH. In this model, the star formation jumps from zero to its maximum value at some time $T_{0}$, after which star formation declines exponentially with some time scale $\tau$, i.e.

\begin{equation} 
SFR (t) \propto
\begin{cases}
\exp( - \frac{t-T_0}{\tau}), & \text t > {T_0} \\
0, & \text t < {T_0}.
\end{cases}
\end{equation}
 \\
Here, $T_{0}$ (Gyr) $\sim$ (0, 15) and $\tau$ (Gyr) $\sim$ (0.3, 10)  (Carnall et al. \citeyear{Carnall2019}).\\

Hence for a pdf, SFR ($t$) = $\lambda\exp(-\lambda t)$, where, $\lambda$ = $\frac{1}{\tau}$ and $T_{0} = 0$. This is the standard form of negative exponential pdf.\\

The exponential model is often used as fiducial model and they are less appropriate at higher redshifts (Reddy et al. \citeyear{Reddy2012}). They have some bias when stellar mass, star formation rate (SFR) and mass-weighted age are reproduced by fitting mock observations of simulated galaxies (Simha et al. \citeyear{Simha2014}, Pacifici et al. \citeyear{Pacifici2016}, Carnall et al. \citeyear{Carnall2019}).

\subsection{Delayed exponentially declining SFR(t)} \label{sec:Delayed exponentiallly declining SFR(t)}
   
Sometimes, delayed exponentially declining SFHs are more realistic than ordinary exponential type. As the stars are formed, they take some time for their evolution from main sequence to giant form and ultimately end their lives for supernovae explosion in case of very massive stars. The ejection of material from supernovae enrich the medium for second generation of the star formation and there is a delay in the recycling of the material (Pagel and Tautvaisiene \citeyear{Pagel1995}). This results in a mere flexible and physical model of the star formation. This shows a rising SFH if $\tau$ is large. Thus,
   
\begin{equation} 
SFR (t) \propto
\begin{cases}
(t-{T_0})\exp( - \frac{t-T_0}{\tau}), & \text t > {T_0} \\
0, & \text t < {T_0},
\end{cases}
\end{equation}   
   
where, $T_{0}$ (Gyr) $\sim$ $(0,15)$, $\tau$ (Gyr) $\sim$ $(0.3,10)$.
   
\subsection{Double power law SFR(t)} \label{sec:double power law SFR(t)}
   
Here the rising and falling slopes of SFH are different over time and are denoted by $\beta$ and $\alpha$ respectively. The function shows a good fit to the red shift evolution of the cosmic star formation rate density (Behroozi et al. \citeyear{Behroozi2013}, Gladders et al. \citeyear{Gladders2013}) as well as it produced good fit to SFHs from simulations (Pacifici et al. \citeyear{Pacifici2016}, Diemer et al. \citeyear{Diemer2017}, Carnall et al. \citeyear{Carnall2019}). Thus,

\begin{equation}
\text{SFR}(t) \propto \left[ \left( \frac{t}{\tau} \right)^{\alpha} + \left( \frac{t}{\tau} \right)^{-\beta} \right]^{-1}
\end{equation}                         

where, $\alpha$ $\sim (0.1,1000)$ is the falling slope, $\beta$ $\sim(0.1,1000)$ is the rising slope and $\tau$ (Gyr) $\sim (0.1,15)$ is related to the peak time.

\subsection{ Gaussian mixture modelling SFR(t)} \label{sec:Gaussian mixture modelling SFR(t)}
   
Gaussian mixture also reflects the episodic nature of star formation in many dwarf galaxies as observed by various authors (Lee et al. \citeyear{Lee2012}, Weisz et al. \citeyear{Weisz2014}). The common form of Gaussian mixture model is,
   
\begin{equation}
SFR(t)=qg_{1}(t)+(1-q)g_{2}(t), \quad \quad  0 < q < 1
\end{equation}

where, $g_{1}$(t) and $g_{2}$(t) are two Gaussian pdfs with parameters ($\mu_{1}$, $\sigma_{1}$) and ($\mu_{2}$, $\sigma_{2}$) respectively and 

\begin{equation}
SFR(t)=q_{1}g_{1}(t) + q_{2}g_{2}(t)+(1-q_{1}-q_{2})g_{3}(t)
\end{equation}

where, $g_{1}$(t), $g_{2}$(t), $g_{3}$(t) are three Gaussian pdfs with parameters ($\mu_{1}$, $\sigma_{1}$), ($\mu_{2}$, $\sigma_{2}$), ($\mu_{3}$, $\sigma_{3}$) in case of mixture of three Gaussian distributions and $q_{1}$, $q_{2}$ are the weights lying between 0 and 1. All the samples are generated using Markov Chain Monte Carlo Method (Robert $\&$ Casella  \citeyear{Robert2004}) where we have used Uniform distribution U[0,1] as the prior distribution.

\subsection{Beta distribution of SFR(t)}
The standard Beta distribution is of the form,
\begin{equation}
    \text{SFR}(t) = \frac{t^ {p-1}(1-t)^ {q-1}}{B( p , q)},  \quad \quad \quad 0 < t < 1
\end{equation}
where $p$, $q$ $>$ 0 are respectively the rising and falling slope of SFH over time and $B(p , q)$ = $\frac{\Gamma(p)\Gamma(q)}{\Gamma(p+q)}$ , where $\Gamma$ is the Gamma function.
Unimodal star formation rate (SFR(t)) has been seen in many giant galaxies (Debsarma et al. \citeyear{Debsarma2016}, Feldmann \citeyear{Feldmann2017}, Das et al. \citeyear{Das2020}) under various parametric conditions. Here we consider the values of $(p, q)$ as (2, 2) and (5, 1) respectively.
   
\section{Luminosity functions (LF)}\label{sec:Luminosity functions}
\subsection{Observed luminosity function of Gaia Catalogue}
   
   In the Gaia Catalogue (\url{https://vizier.u-strasbg.fr/viz-bin/VizieR-3?-source=I/345/gaia2}), from the first table (\url{https://vizier.u-strasbg.fr/viz-bin/VizieR-3?-source=I/345/gaia2}) of 169,29,19,135 observations, few parameters like $G_{mag}$ (first one in the 4th block) (here written as $m_{G}$), Plx (10th one in first block) and AG (9th one in the 6th block) are considered, where $m_G$, Plx and AG are the apparent magnitudes, parallaxes and extinction parameters in G band. The sample is restricted to Plx $<$ 20$^{\prime\prime}$, to avoid errors to large extent. This reduces the sample size from 169,29,19,135 to 69,92,398. Thus we obtain 69,92,398 observations from the set (for those whose AG values exist and have positive parallaxes). The observed apparent magnitude $m_{G}$ (in G band) from Gaia dataset are transformed to extinction corrected absolute magnitude $(M_{G})_{0}$ using Pogson relations,
   
\begin{equation}
(M_{G})_{0}= m_{G} - [5\log_{10}\left(\frac{1000}{plx}\right)-5+AG]. 
\end{equation}

Then these  $(M_{G})_{0}$'s are converted to luminosity by,

\begin{equation}
(M_{G})_{0} - (M_{G,\odot})_{0} = -  2.5\log_{10}(L/{L_{\odot}}).
\end{equation}

Luminosities, thus obtained are fitted to a mixture of Normal Distribution of the form,

\begin{equation}
f(l)=\lambda_1N(l;\mu_1,\sigma_{1}^2)+(1-\lambda_1)N(l;\mu_2,\sigma_{2}^2)
\end{equation}
 
\begin{figure}
\centering
\includegraphics[scale=1.6]{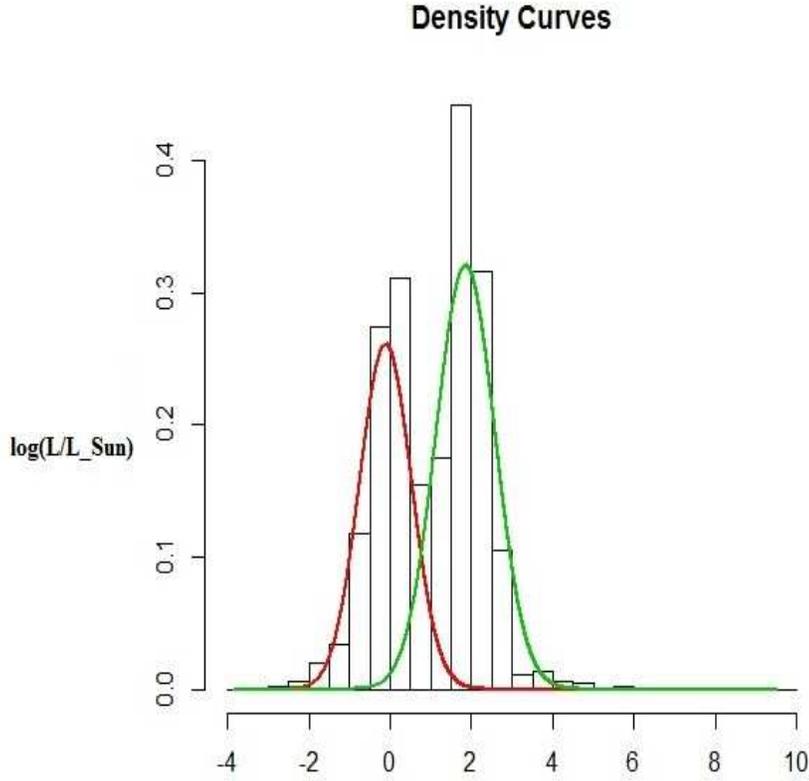}
\caption{Luminosity function fitted to reduced Gaia Mission data set.} \label{ fig:2}
\end{figure}

The maximum-likelihood estimates of the above distribution came out to be,
 $\hat{\lambda_{1}}=0.582439$, $\hat{\mu_{1}}=0.01337392$, $\hat{\mu_{2}}=4.93331006$, $\hat{\sigma_{1}}=1.810862$, $\hat{\sigma_{2}}=1.595570$ respectively, with p-value 0.2216. Fig. \ref{ fig:2} shows the luminosity values with the fitted curves.

\subsection{Truncated power law}  \label{sec:Trucated power law}
 
There are various observations on the luminosity function of young massive clusters. For example, Arches cluster (Figer et al. \citeyear{Figer1999}), NGC 3603 (Sung and Bessell \citeyear{Sung2004}, Stolte et al. \citeyear{Stolte2006}, Harayama et al. \citeyear{Harayama2008}, Zwart et al. \citeyear{Zwart2010}), Westerlund 1 (Brandner et al. \citeyear{Brandner2008}) which are of  truncated power law form. So we have also considered truncated power law for a typical luminosity function. This has the form,
 
 \begin{equation}
 f(l)=\frac{\alpha\beta^{\alpha}l^{-\alpha-1}}{1-(\frac{\beta}{\nu})^\alpha}, \quad \quad \beta<l<\nu 
 \end{equation}
 
 where $\beta$, $\nu$ are the lower and upper values of the luminosities in the isochrones, used in the model (i.e. $\log_{10}(L/{L_{\odot}})$ $\sim$ -2 to +4, $\alpha \sim 1.05$).\\ 
Whitmore et al. (\citeyear{Whitmore2014}) observed 20 nearby (4$-$30 Mpc) star clusters in star forming galaxies based on ACS source lists generated by Hubble Legacy Archive. A typical cluster luminosity function is fitted by a power law pdf of the form,
 
\begin{equation}
f(l)=\lambda\exp(-\lambda l),  \quad \quad l \ge 0 
\end{equation}

where $\lambda$  varies  from $1.95\pm0.02$ to $-3.09\pm0.46$. All the galaxies in the catalogue are of Sa type and their absolute magnitudes $(M_{B})$ in B$-$band vary from $-17.46$ to $-20.96$. We have considered a few values of $\lambda$ in the simulation study to investigate the effect of luminosity functions in external galaxies.

\section{Matching criteria between simulated and observed stellar population distributions}\label{sec:5}

\subsection{Choice of dissimilarity measure}\label{Choice of dissimilarity measure}

Here the problem is to compare two histograms corresponding to observed and simulated distributions. We can use a good dissimilarity measure as this can be used as an inverse matching criteria. More the value of the measure, lesser the degree of similarity between the two distributions. So, for a good measure, the value should be significantly lower while matching two different types of distributions and close to zero for two very similar distributions.
\\
It is well known that Chi-square distance function is a very good dissimilarity measure (Pele and Werman \citeyear{Pele2010}, Yang et al. \citeyear{Yang2015}, Greenacre \citeyear{Greenacre2017}) for comparing two histograms.\\
If  $p_{j}$ and  $q_{j}$ be the probabilities (frequency densities) corresponding to $j^{th}$ class $(j = 1,2,...,k)$ of the two histograms under consideration then the Chi-square distance function is given by,

\begin{equation}
\chi^2_{d} = \frac{1}{2} \sum_{j = 1}^{k} \frac{(p_j - q_j)^2}{(p_j + q_j)},
\end{equation}
i.e. $p_j$ and $q_j$ correspond to the bin values of $j^{th}$ class with respect to the two histograms under consideration. $\chi^2_{d}$ satisfies all the necessary properties of a dissimilarity measure. 
      
\subsection{Tuning of optimum bin size}\label{Tuning of optimum bin size}

While drawing the histogram it is necessary to choice the bin size optimally by taking into consideration the dissimilarity measure and ranges of the variables under consideration. In order to study the discriminating power of the dissimilarity measure and to choose optimum bin size, we have defined the following metric. Primarily we have chosen one simulated data (model) and  some similar and dissimilar data sets (observed) through physical investigation.\\
Then the metric under consideration is,\\

 metric = (mean distance between simulated data set and dissimilar data set) $-$ (mean distance between simulated data set and similar data set).\\
 
Here the mean is taken over dissimilar data sets and similar data sets respectively in first and second term of the metric. \\
More the value of the metric (i.e. difference) for a particular dissimilarity measure, the larger is its discrimination power.
One thing to note that the bin size also depends on the ranges of the histogram. At the time of similarity measure between one model dataset and one observed dataset while calculating the minimum and maximum ranges of the variables we include the variables of model and observed datasets with the anchor, similar and dissimilar datasets. With the calculated range values we did optimum bin size tuning on the anchor, similar and dissimilar dataset (but with only Chi-Square distance) as described above and calculated the optimum bin size for which the metric value becomes largest. Finally with the optimum bin size value we calculate normalized histograms for both model and observed dataset and finally calculate the similarity/dissimilarity with Chi-Square distance. The whole algorithm is tabulated in Algorithm \ref{algorithm}. From the given Table \ref{tab:1}, it can be seen that while calculating the validation metric for the mentioned Observed Data and Model Data, we find that for bin size 50, the metric is largest with value 0.5074. So, 50 is the optimum bin size while calculating the similarity between the mentioned two data.

\begin{table}[h]
\centering
\begin{tabular}{|c|c|}
\hline
\textbf{Bin Size} & \textbf{Metric on Chi-Square Distance} \\ \hline
10 &  0.5067 \\ \hline
20 & 0.4881 \\ \hline
30 &  0.4864 \\ \hline
40 & 0.5000 \\ \hline
50 &  0.5074 \\ \hline
60 & 0.5021 \\ \hline
70 &  0.4962 \\ \hline
80 &  0.4870 \\ \hline
90 &  0.4985 \\ \hline
100 & 0.4852 \\ \hline
\end{tabular}
\caption{Values of metric on Chi-Squared distance between observed data: H and $ \chi$ Persei (an open cluster) and model data  for Gaussian SFR(t) with three modes (means $\sim$ 9, 11, 13 and sds $\sim$ 0.05, 0.05, 0.05) and Gaia LF for different bin sizes.} \label{tab:1}
\end{table}

\begin{algorithm}
\caption{Optimum Bin Size Selection Algorithm}\label{algorithm}
 \begin{algorithmic}[1]
    \State Calculate the minimum and maximum of the ranges of \textbf{color} variable from the validation datasets (anchor, similar and dissimilar datasets), model dataset and observed dataset.
    \State Calculate the minimum and maximum of the ranges of \textbf{log (L/$L_{\odot}$)} variable from the validation datasets (anchor, similar and dissimilar datasets), model dataset and observed dataset.
      \For{$bin = 10, 20,\ldots, N$}
        \State Calculate \textbf{metric} from the above \textbf{metric} calculation equation using Chi$-$Squared distance.
      \EndFor
    \State Find the bin size for which metric value is maximum.
    \State Build normalized histograms for both model and observed data.
    \State Calculate Chi$-$Squared dissimilarity measure for the above two histograms.
\end{algorithmic}
\end{algorithm}

\subsection {Actual data analysis}\label{Actual data analysis}
  
Here our target is to compare several observed CMDs with possible simulated CMDs generated under different model assumptions. As corresponding to each CMD we have data for two variables viz. `colour' and 'luminosity' (viz. log (L/$L_\odot$)), we have to draw one normalized bivariate histogram corresponding to each CMD. As shown in Table \ref{tab:1}, the optimum bin size may be taken as 50 $\times$ 50 for each such bivariate histogram as the ranges of each of the two variables are almost similar for all the data sets under consideration (both in the present subsection \ref{Choice of dissimilarity measure} and subsection \ref{Tuning of optimum bin size}).\\
Then we have computed the Chi-square dissimilarity measure values for different pairs of simulated (model) and observed CMDs in order to find out the possible matchings. The simulated CMDs i.e. synthetic H$-$R diagrams have been generated using Monte Carlo simulation under various choices of the concerned parameters.

\begin{figure}[h]
    \centering
    \includegraphics[scale=0.9]{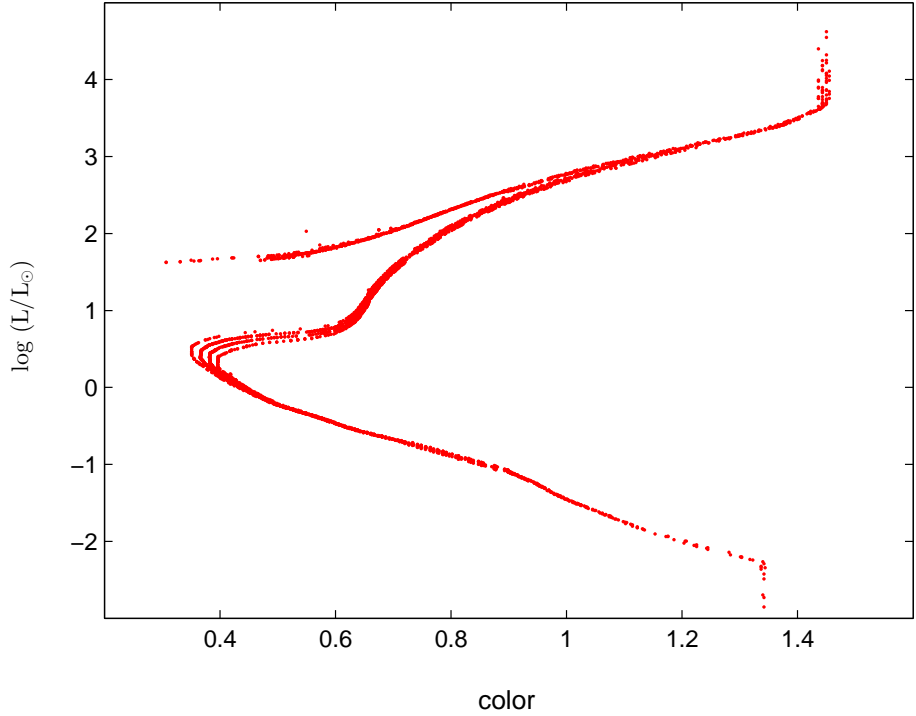}
    \caption{  Synthetic H$-$R diagram for bimodal Gaussian (2 modes) type of SFH (i.e. $SFR(t)$) ($\mu_{1}$ = 11, $\sigma_{1}$ = 0.2, $\mu_{2}$ = 13, $\sigma_{2}$ = 0.5, q = 0.5), LF = Gaia LF and $ Z$ = 0.0004.} \label{fig:3}
\end{figure}

\begin{figure}[h]
    \centering
    \includegraphics[scale=0.9]{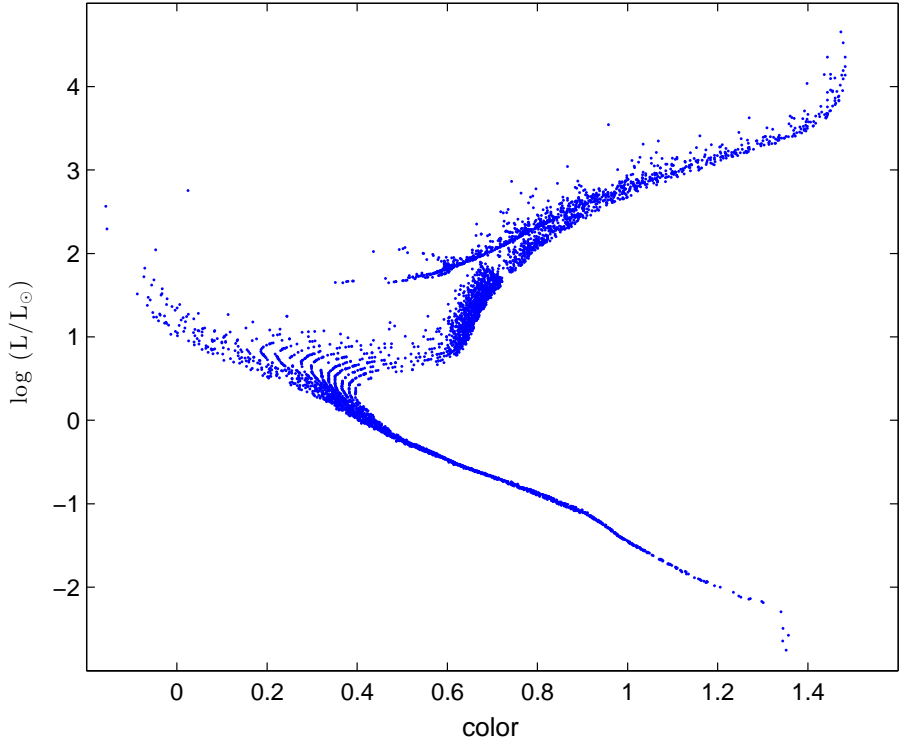}
    \caption{ Synthetic H$-$R Diagram for Beta distribution of SFH (i.e. SFR(t)) with parameters $p$= 2, $q$= 2, LF = Gaia LF and $Z$ = 0.0004 } \label{fig:4}
\end{figure}

\begin{figure}
    \centering
    \includegraphics[scale=1]{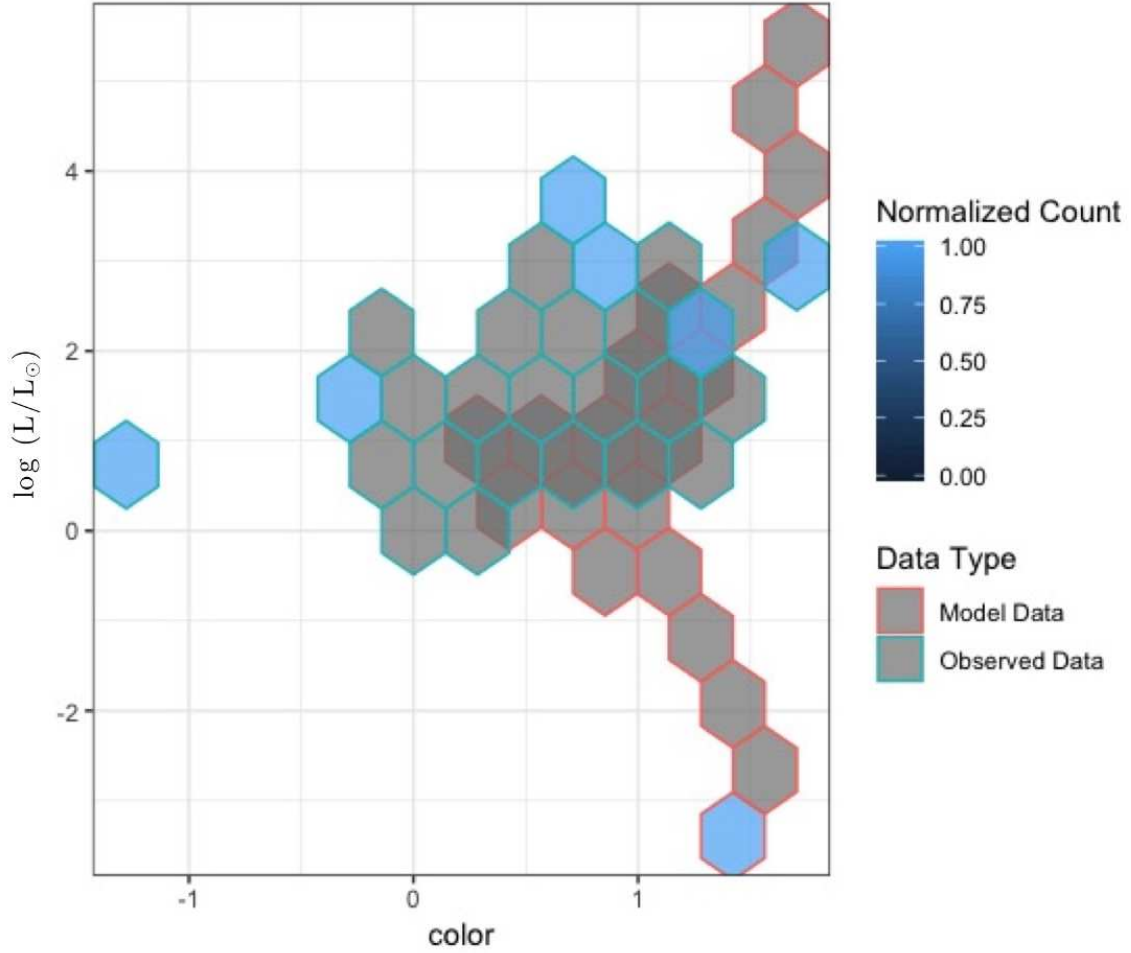}
    \caption{(a) Comparison of Normalized Histograms and (b) Scatter Plots between Observed Data: SMC Br\"{u}ck 2 ( an SMC star cluster) and  Model Data:  with Gaussian (3 modes) $SFR(t)$ (means $\sim$ 2, 5, 7 and sds $\sim$ 0.05, 0.05, 0.05) and Gaia LF and $Z$ = 0.0004.  Here bin size is 10 and minimum Chi-squared dissimilarity value is 0.265.}
    \label{fig:5}
\end{figure}

\begin{figure}
    \centering
    \includegraphics[scale=1]{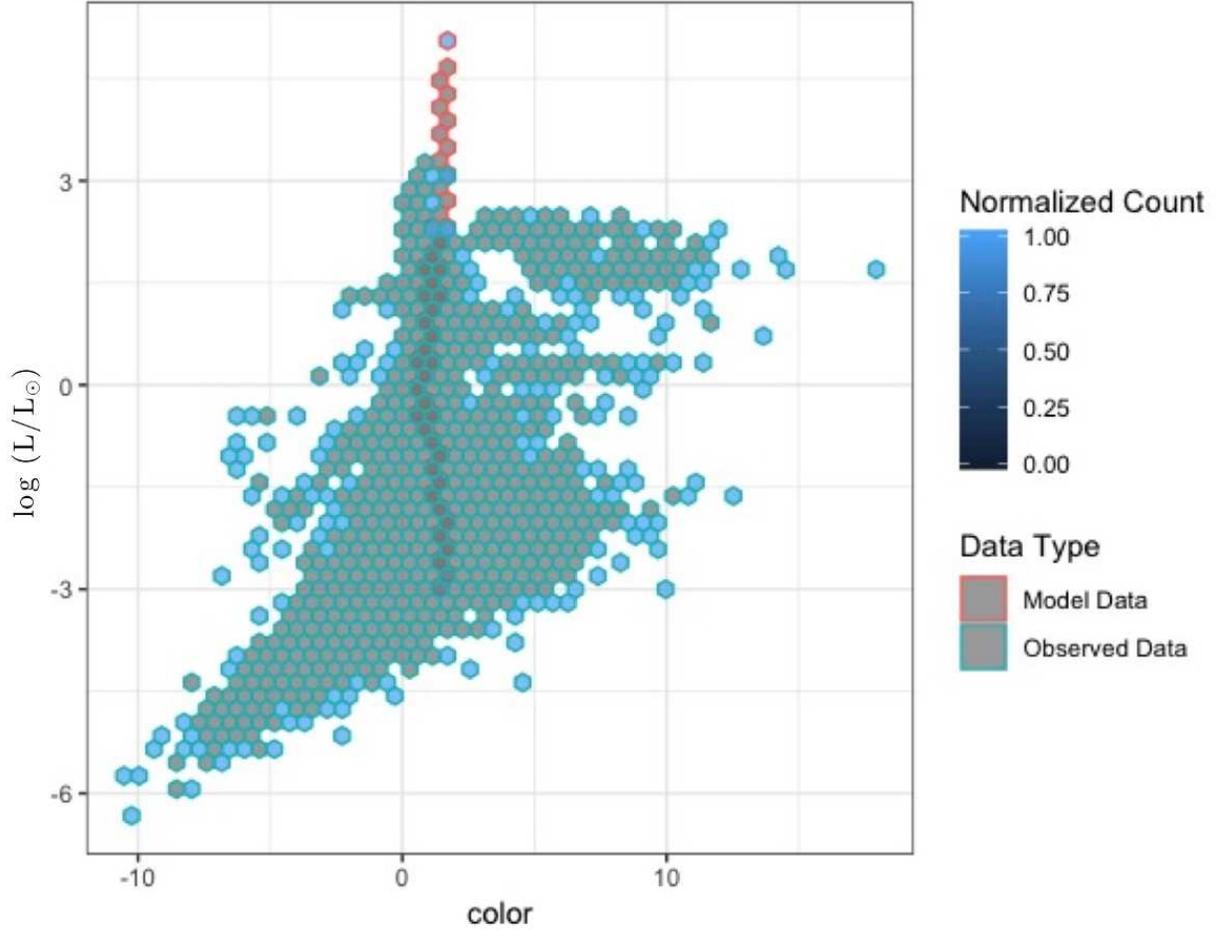}
    \caption{(a) Comparison of Normalized Histograms and (b) Scatter Plots between Observed Data: H and $\chi$ Persei ( an open MW cluster)  and Model Data with $SFR(t)$ Gaussian (2 modes) type (means $\sim$ 2, 8 and sds $\sim$ 0.2, 0.5 respectively), Gaia LF and $Z$ = 0.0004. Here bin size is 50 and Chi-squared dissimilarity value is 0.577.}\label{fig:6}
\end{figure}

\begin{figure}
    \centering
    \includegraphics[scale=1.1]{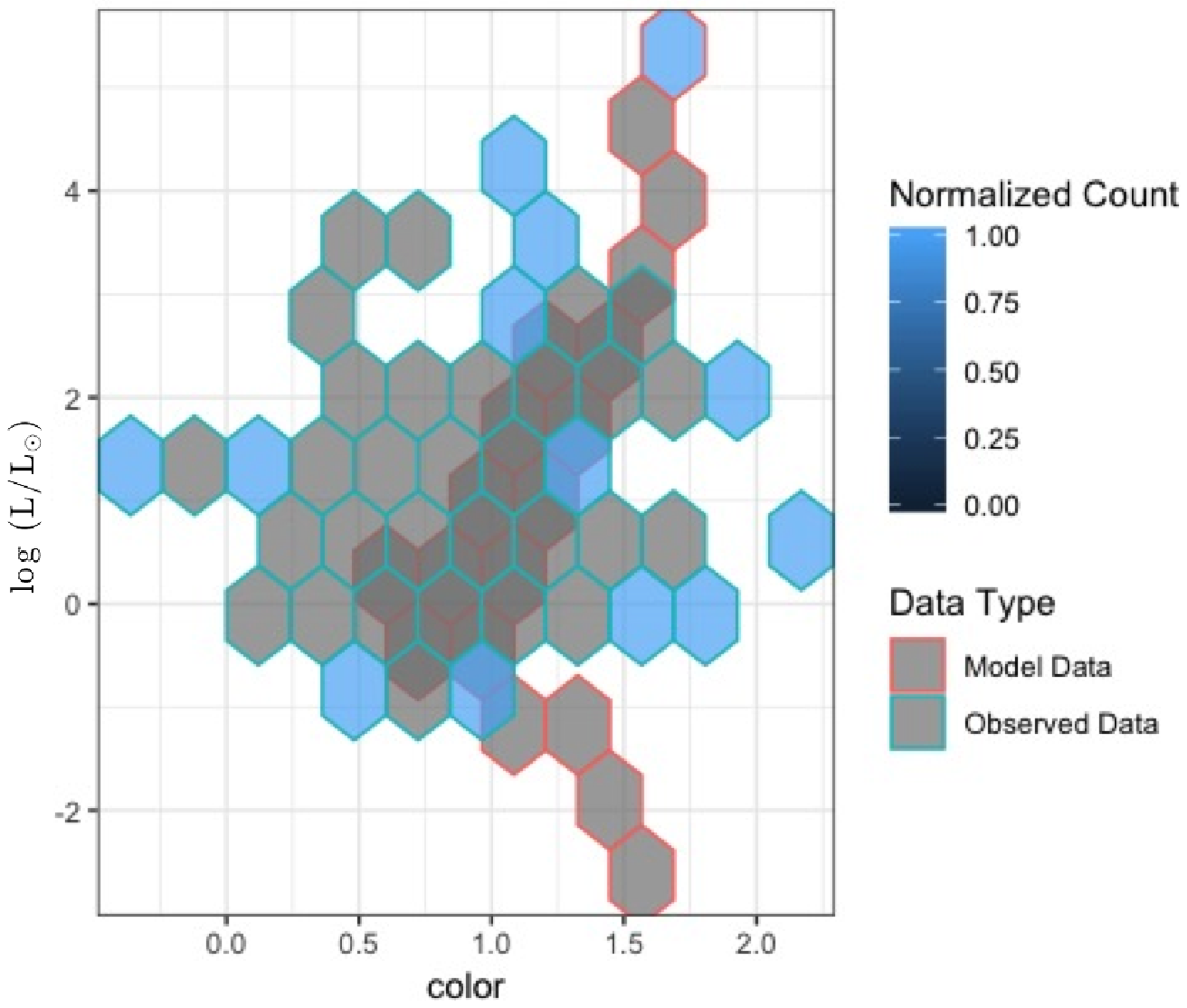}
    \caption{(a) Comparison of Normalized Histograms and (b) Scatter Plots between Observed Data: 47 Tuc (MW GC) and Model Data with $SFR(t)$ of Gaussian (3 modes) type (means $\sim$ 2, 5, 7 and sds $\sim$ 0.05, 0.05, 0.05) and Gaia LF and $Z$ = 0.019.  Here bin size 10 and chi-squared similarity value of 0.086.}\label{fig:7}
\end{figure}

 \section{Results and discussions} \label{Results and discussions}
 \subsection{Properties of the synthetic H-R diagrams}\label{Properties of the synthetic H-R diagrams}
 
 In the present problem, synthetic H-R diagrams have been generated using Monte Carlo simulation under various star formation histories, luminosity profiles and metallicities and the diagrams have been compared to various star clusters, both Galactic or globular clusters, to explore the origin of these star clusters. The matching is carried out through an algorithm of finding the minimum distance between any pair of choice, viz. synthetic H-R diagram and observed star cluster. The details of the procedure has been outlined in Section \ref{sec:5}. Finally, the pair with minimum distance has been chosen as the pair for comparison. The values of the dissimilarity measures having values $<$ $1.0$, have been listed in various tables (Tables \ref{tab:2}$-$\ref{tab:16}). The properties of GCs (metallicities and Galacto-centric distances etc.) are from Harris (\citeyear{Harris1996}) updated catalogue (listed in the appendix (Table \ref{tab:17})).The following observations have been reflected from the tables.\\
 
\noindent Table \ref{tab:2}:
 
\begin{enumerate}[(i)]

    \item Milky way globular clusters (GCs) having higher metallicity $( [Fe/H] \sim-0.70)$, closer to the centre are formed primarily with Gaia LF and SFH (SFR(t)) of Gaussian mixtures with either three ($\sim$ 2, 5, 7 Gyrs) or two ($\sim$ 2, 8 Gyrs) modes.
    
    \item Milky way GCs with lowest metallicity $( [Fe/H] \sim-2.28)$, closer to the Galactic centre form with a Gaia LF and SFH of Gaussian mixture with two modes.
    
    \item Milky Way GCs having lower metallicity ($ [Fe/H] \sim$ $-$2.10 to $-$1.27) and far from the Galactic centre ($>$ 40 kpc) are formed with Gaia LF and SFH of Gaussian mixture type with three or two modes and GCs farthest from the Galactic centre ($\sim >$ 100 kpc) with very low metallicity ($[Fe/H] < -2.0$) have Gaussian SFH with two modes.
    
    \item Milky Way GCs with very high metallicity $( [Fe/H] \sim-0.50$ to $-$0.20), and close to the Galactic centre have been formed with SFH of Gaussian mixture distribution along with Gaia LF.
    
\end{enumerate}
 
\noindent Table \ref{tab:3}:

\begin{enumerate}[(i)]
    \item All Small Magellanic Cloud clusters could be originated with star formation history (SFR(t)) of Gaussian mixture distributions with three  modes at 2, 5 and 7 Gyrs  along with Gaia LF.
\end{enumerate}
 
\noindent Table \ref{tab:4}:

\begin{enumerate}[(i)]

    \item  All open clusters in our Galaxy could be originated with star formation history (SFR(t)) of Gaussian mixture distributions with three or two modes but in particular $\sigma$ Orionis remains independent of the nature of star formation history.
\end{enumerate}
 
So, from the above observations it is clear that in case of stellar populations, Gaussian LF (Gaia in the present case)  associated with Gaussian SFH, can lead to the formation of star clusters.\\
 
When the luminosity profile changes from Gaussian to exponential one or truncated power law the following properties of the H$-$R diagram are being observed.\\
 
\noindent Table \ref{tab:5}:

\begin{enumerate}[(i)]

    \item For Milky Way GCs with high metallicity $( [Fe/H] \sim -0.70)$ and close to the Galactic centre ($< 10$ kpc), may originate with star formation history of Gaussian mixture nature of two modes along with exponential distribution of luminosity with steep slope ($\lambda$ high) in their H$-$R diagram.
    
    \item  Milky Way GCs having intermediate metallicity ($[Fe/H]< -1$) and  closer ($<$ 10 kpc) to the Galactic centre may result either with exponential star formation history ($\lambda\sim10)$ and truncated power law luminosity or with truncated power law luminosity and double power law star formation history.
    
    \item Milky Way GCs with very low metallicity $( [Fe/H] < -2.0)$, and far from the Galactic centre may result with double power law SFH and truncated power law LF.
    
    \item Milky Way GCs with very high metallicity $( [Fe/H] \sim-0.50)$, closer to the Galactic centre may result with truncated power law luminosity and double power law SFH.
    
    \item Milky Way GCs having high metallicity $( [Fe/H] \sim-0.50)$ and farthest from the Galactic centre may also result with double power law SFH and truncated power law LF.
    
    \item Milky Way GCs, closest to the Galactic Centre ($\sim 5$ kpc) having intermediate metallicity $([Fe/H] \sim -1.95)$ may result with double power law SFH and exponential or truncated power law LFs.

\end{enumerate}

\noindent Table \ref{tab:6}:

\begin{enumerate}[(i)]

    \item  All SMC star clusters may result in most situations with truncated power law LF and double power law SFH.
    
\end{enumerate}

\noindent Table \ref{tab:7}:

\begin{enumerate}[(i)]

    \item  All Milky Way open clusters may originate with truncated power law LF and double power law SFH.
    
\end{enumerate}

All the properties mentioned above were simulated for Z $= 0.0004$ through SSP model.\\
  
\noindent Table \ref{tab:8}:
 
\begin{enumerate}[(i)]

     \item The H$-$R diagrams of Milky Way GCs with high metallicity ($ [Fe/H] \sim -0.76$) and closer to the Galactic centre may originate with Gaia LF, double power law SFH and for Z= $0.019$.
     
     \item Milky Way GCs with intermediate metallicity ($ [Fe/H] <-1.20$) closest ($<7.5$ kpc) to the Galactic centre may originate with Gaia LF and double power law SFH. GCs with intermediate metallicity ($ [Fe/H] \sim-1.14$) and still closer to the Galactic centre ($< 9.6$ kpc) may originate with Gaia LF and double power law SFH for Z $= 0.019$.
     
     \item Milky Way GCs with very low metallicity ($ [Fe/H] \sim -2.28$) and closer to the Galactic centre may originate with Gaia LF and  SFH of double power law or Beta distribution for Z= 0.019.
     
     \item  Milky Way GCs with still lower metallicity ($ [Fe/H]\sim -2.12$) and farthest ($\sim 100$ kpc) from the Galactic Centre may originate with Gaia LF and SFH of Beta type or double power law.
     
     \item Milky Way GCs with intermediate metallicity ($ [Fe/H]\sim -1.83$) and far from the Galactic Centre ($\sim 21.3$ kpc) may originate with Gaia LF and SFH of Beta type.
     
     \item Milky Way GCs closer to the Galactic Centre ($<$ 10 kpc) having  high metallicity ($ [Fe/H]\sim -0.70$) may be originated with Gaia LF and SFH of one of Gaussian type with three modes or of double power law type.
     
     \item Milky Way GCs far ($<$ 10 kpc ) from the Galactic centre having high metallicity ($ [Fe/H]\sim -0.73$) may originate with Gaia LF and Gaussian mixture (3 modes; 2, 5, 7 Gyr) type SFH for Z $= 0.019$.
     
 \end{enumerate}

\noindent Table \ref{tab:9}:

\begin{enumerate}[(i)]

    \item All clusters in SMC may originate with Gaia LF and star formation history (SFH) of Beta type for Z = 0.019.
    
\end{enumerate}

\noindent Table \ref{tab:10}:

\begin{enumerate}[(i)]

    \item All open clusters in our Galaxy may originate with Gaia LF and double power law star formation history (SFH) for Z = 0.019.
    
\end{enumerate}

\noindent Tables \ref{tab:11} - \ref{tab:13}:

\begin{enumerate}[(i)]

    \item  From tables \ref{tab:11} - \ref{tab:13}, it is clear that in almost all cases the dissimilarity measures are close to $\sim 1.0$ and there is no large differences in values for exponential luminosity for Z = 0.019. So, in a metal rich medium, a combination of exponential luminosity  along with different star formation history is not a good choice for the origin of formation of Milky Way GCs or open clusters or SMC star clusters.
    
\end{enumerate}
  
\noindent Tables \ref{tab:14} - \ref{tab:16}:

\begin{enumerate}[(i)]

    \item From tables \ref{tab:14} - \ref{tab:16}, it is also clear that in a metal rich medium the truncated power law luminosity with various types of star formation histories cannot lead to the formation of  of star clusters we generally observe in our Galaxy or in external galaxies.
    
\end{enumerate}

\subsection{Origin of various star clusters explored}

In the previous section we have found similarities of various model CMDs generated using various input parameters e.g. star formation history, luminosity function, heavy element abundance with observed CMDs of various star clusters. The following interpretations might lead to explain the origin of the formation of those observed clusters.

\begin{itemize}

    \item It is found that for low metallicity medium (Z = 0.0004), globular clusters having higher metallicity and closer to the centre or farthest from the Galactic centre have multiple modes of the distribution of SFH (e.g. Gaussian mixture model with 3 modes) with Gaia luminosity function. Due to several modes, star formation is of episodic nature and stars which are formed in first generation enriched the medium for the formation of second generation stars and so on. As generations proceed, the metallicity becomes higher (here $[Fe/H]$) and GCs having higher metallicity and closer to the centre belong to inner halo populations (Burkert $\&$ Smith \citeyear{Burkert1997}, Mackey $\&$ van der Bergh \citeyear{Mackey2005}, Chattopadhyay $\&$ Chattopadhyay \citeyear{Chattopadhyay2007}, Dotter et al. \citeyear{Dotter2011}, Lamers et al. \citeyear{Lamers2017}).
    
    \item GCs closer to the centre but having minimum metallicity have not gone through several episodes of star formation to enhance their metallicity, but due to Gaia type luminosity function they have few massive stars (e.g. observations (i), (ii) and (iii), from Table \ref{tab:2})(viz. subsection \ref{Properties of the synthetic H-R diagrams}).
  
  \item GCs having intermediate type metallicity have not gone through episodic star formation type history but through an exponential type of star formation history accompanied by an exponential type of luminosity function. High value of the constant $\lambda$ (e.g. 10) corresponds to steeper growth rate in a very short interval of time which inhibits growth of metallicity to a very large extent (Table \ref{tab:5}, observation (ii)).
  
  \item GCs farthest from the Galactic centre are accreted from satellite galaxies having their velocity dispersion very high (Mackey $\&$ Gilmore \citeyear{Mackey2004}, Chattopadhyay $\&$ Chattopadhyay \citeyear{Chattopadhyay2007}, Mondal et al. \citeyear{Mondal2008}, Duncan $\&$ Terry \citeyear{Duncan2010}, Dotter et al. \citeyear{Dotter2011}, Keller et al. \citeyear{Keller2012}, Veljanoski et al. \citeyear{Veljanoski2014}). So their metallicities are very poor which may occur due to smaller number of episodes of star formation (Table \ref{tab:5}, observation (iii)) and truncated power law LF.
  
  \item GCs having higher metallicity may result from other SFH having double power law distribution and truncated or exponential power law luminosity. Now for a truncated or exponential power law type luminosity, number of massive stars are lower than those of less massive stars. In that situation, less massive stars may coagulate to enhance the metallicity
  (Mackey $\&$ Gilmore \citeyear{Mackey2004}, Beasley et al. \citeyear{Beasley2008},
  Lamers et al. \citeyear{Lamers2017})
  (Table \ref{tab:5}, observations (iv), (v)).
  
  \item Milky Way open clusters and SMC clusters have similar mechanism of star formation i.e. exponentially luminosity function and Gaussian or double power law SFH. \\
  \newline
  When the star formation occurs in an initially enriched medium (e.g. Z = 0.019), the following interpretations are possible.
  
  \item In an enriched medium of heavy elements, GCs close to the Galactic centre may result from double power law SFH distribution (i.e. SFR(t)) which is unimodal for Gaia type luminosity function. On the contrary, when SFH distribution is a Gaussian mixture with lesser number of modes then exponential distribution of luminosity function with steep slope is preferred. These results can be interpreted in the manner that for Gaia luminosity there is a large fraction of massive stars of higher metallicity. So, a double power law type SFH may lead to even higher metallicity. This reflects the combined effect of enriched medium and presence of large fraction of massive stars producing heavy metals. The opposite behaviour i.e. exponential or Beta type luminosity function with Gaussian mixture model type SFH needs more episodic star formation events to attain the said metallicity (Table \ref{tab:8}, observations (i)$-$(vi)).
  
  \item Similar logic of first kind is applicable to SMC clusters and open clusters.
  
\end{itemize}

On the basis of the above mentioned study we can at least say that formation of GCs depends on various factors like SFH ( SFR(t)), LF and heavy element abundance (Z) of the star forming region. Same GCs may originate from various SFH or LF provided the heavy element abundances of the region are different. Hence, SFH and LF should be thoroughly studied to find the origin of any star cluster. The above study motivates further analysis along this direction.

\section{Conclusion}\label{Conclusion}
 
In the present study we have generated synthetic H$-$R diagrams (here CMD) of various model star clusters with wide variety of SFH (or SFR(t)), luminosity function and heavy element abundance of the star forming region. The corresponding bivariate histograms have been compared with respect to Chi-square distance criterion. \\

The following facts have been uncovered in the above study.
 
\begin{itemize}

    \item We have considered the H-R diagrams of a large number of GCs, open clusters and SMC Clusters.
    
    \item The synthetic H-R diagrams are simulated with respect to SFH, LF and Z. This is a new approach to explore the origin of the formation of several star clusters.
    
    \item The matching procedure includes distance computations on the basis of two dimensional histograms of the colour-magnitude diagram. This is an innovative approach for dealing with such type of problems in Astrophysics.
    
    \item The choice of optimum bin size is done in an objective manner.
    
    \item It is found that for a poorly enriched medium Gaussian mixture model of SFH along with Gaussian LF (e.g. Gaia luminosity function) may result in GCs. A combination of Gaussian LF with exponential or double power law SFR(t) may result to the formation of GCs in highly enriched medium.
    
    \item For SMC and open clusters both the media ( i.e. poor or highly enriched) may form star clusters. A combination of Gaia LF and Beta type or double power law (DPL) SFH are responsible in highly enriched medium whereas Gaia LF with Gaussian SFH are responsible in poorly enriched medium.
    
\end{itemize}

Since the formation of star clusters is a very complicated method starting from fragmentation of a gravitationally unstable big cloud and subsequent evolution through coalescence, accretion and tidal evaporation, the above study will motivate researchers in several directions.

\begin{table}
\centering
\renewcommand{\arraystretch}{0.72}
\caption{Comparison of synthetic H-R diagrams using Gaia LF, Z = 0.0004 with observed MW, GCs for various SFR(t). DPL, $\beta$, DE, E, G3, G2 stand for Double Power Law, Beta Distribution, Delayed Exponential, Exponential, Gaussian mixture model with 3 modes and Gaussian mixture model with 2 modes for various SFR(t) respectively.}\label{tab:2}
\resizebox{\textwidth}{!}{
\begin{tabular}{l|c|c|c|c|c|c|c|c|c|c|c}
\hline
Names & DPL & DPL & $\beta$ & $\beta$ & DE & E & E & G3 & G3 & G2 & G2 \\
of & ($\alpha$, $\beta$) & ($\alpha$, $\beta$) & (p, q)  & (p, q)  & ($T_{0}$, $\tau$) & ($\lambda$) & ($\lambda$) & ($\mu_{1}$, $\sigma_{1}$) & ($\mu_{1}$, $\sigma_{1}$) &  ($\mu_{1}$, $\sigma_{1}$) & ($\mu_{1}$, $\sigma_{1}$) \\ 
observed & $(\tau$) & ($\tau$) &  &  &  &  &  & ($\mu_{2}$, $\sigma_{2}$)  & ($\mu_{2}$, $\sigma_{2}$) & ($\mu_{2}$, $\sigma_{2}$) & ($\mu_{2}$, $\sigma_{2}$) \\
star & (50, 0.2) & (50, 10)  & (2, 2) & (5, 1) & (15, 9) & (5) & (10) & ($\mu_{3}$, $\sigma_{3}$) & ($\mu_{3}$, $\sigma_{3}$) & (q) & (q) \\
clusters & (15) & (15) &  &  &  &  &  & ($q_{1}$, $q_{2}$) & ($q_{1}$, $q_{2}$)  &  &  \\
 &  &   &  &  &  &  &  & (9, 0.05) & (2, 0.05) & (2, 0.2) &  (11, 0.2) \\
 &  &  &  &  &  &  &  & (11, 0.05) & (5, 0.05) & (8, 0.5) & (13, 0.5) \\
  &  &  &  &  &  &  &  & (13, 0.05) & (7, 0.05) & (0.5) & (0.5)\\
&  &  &  &  &  &  &  & (0.5, 0.3) & (0.5, 0.3) &  &  \\
\hline
47 Tucanae  & 0.4600 & 0.4525 & 0.4573 & 0.4562 &	0.4748 & 0.5857 & 0.5881 & 0.4557 & 0.1901 & 0.1238	& 0.450  \\
(Alcaino \citeyear{Alcaino1987}) &  &  & &  &	 &  &  &  &  &	&  \\
\hline
M 14 & 0.7743 & 0.7738 & 0.7556 & 0.7730 & 0.7599 & 0.8791 & 0.9102 & 0.7729 & 0.2883 & 0.3204 & 0.7730 \\
(Contreras \citeyear{Contreras2013}) &  &  & &  &	 &  &  &  &  &	&  \\
\hline
M 5 & 0.3143 & 0.3136 &	0.3130 & 0.3144 & 0.3319 & 0.4411 &	0.4428 & 0.3189 & 0.1182 & 0.1257 & 0.3089 \\
(Sandquist \citeyear{Sandquist1996}) &  &  & &  &	 &  &  &  &  &	&  \\
\hline
M 92 &	0.3402 & 0.3503 & 0.3411 & 0.3422 &	0.3431	& 0.4415 & 0.4421 & 0.3468 & 0.2626 & 0.2354 & 0.3451 \\
(Stetson \citeyear{Stetson1988}) &  &  & &  &	 &  &  &  &  &	&  \\
\hline
M 53 &	0.7976 & 0.8202 & 0.7918 &	0.8285 &	0.8079 &	0.7787 &	0.7939 &	0.8223 &	0.5366 &	0.5396 &	0.8270 \\
(Heasley \citeyear{Heasley1991}) &  &  &  &  &	 &  &  &  &  &	&  \\
\hline
NGC 1261 &	0.6196 &	0.5895 &	0.5755 &	0.5762 &	0.5834 &	0.6749 &	0.6982 &	0.5830 &	0.3681	& 0.4407 &	0.5795 \\
(Alcaino \citeyear{Alcaino1992}) &  &  &  &  &	 &  &  &  &  &	&  \\
\hline
NGC 1904 & 	0.8791 &	0.9972 &	0.9972 &	0.9972 &	0.9972 &	0.9972 &	0.8788 &	0.9972 &	0.8789 &	0.8795 &	0.9972 \\
(Kravtsov \citeyear{Kravtsov1997}) &  &  &  &  &	 &  &  &  &  &	&  \\
\hline
NGC 2808 & 0.6506 &	0.6605 &	0.6413 &	0.6502 &	0.6506 &	0.7958 &	0.7924 &	0.6585 &	0.2276 &	0.2298 &	0.6492 \\
(Sosin \citeyear{Sosin1997}) &  &  &  &  &	 &  &  &  &  &	&  \\
\hline
NGC 288 &	0.6175 &	0.6291 &	0.6147 &	0.6172 &	0.6311 &	0.7142 & 	0.7160 &	0.6310 &	0.3609 &	0.2985 &	0.6162 \\
(Alcaino \citeyear{Alcaino1997}) &  &  &  &  &	 &  &  &  &  &	&  \\
\hline
NGC 3201 & 0.7456 &	0.7282 & 0.7217 & 0.7281 & 0.7426 &	0.8494 & 0.8442 & 0.7361 & 0.2428 &	0.2762 & 0.7253 \\
(Layden \citeyear{Layden2003}) &  &  &  &  &	 &  &  &  &  &	&  \\
\hline
NGC 362 & 0.6795 & 0.6654 &	0.6624 & 0.6527 & 0.6699 & 0.8565 & 0.8589 & 0.6679 & 0.3940 &	0.3304 & 0.6530\\
(Green \citeyear{Green1990}) &  &  &  &  &	 &  &  &  &  &	&  \\
\hline
NGC 4147 & 0.5603 &	0.5268 & 0.5264 & 0.5152 & 0.5395 & 0.6533 & 0.6751 & 0.5286 & 0.4323 &	0.3569 & 0.5159 \\
(Wang \citeyear{Wang2000}) &  &  &  &  &	 &  &  &  &  &	&  \\
\hline
NGC 4372 & 0.8031 &	0.7998 & 0.7995	& 0.7996 & 0.8074 &	0.9358 & 0.9324 & 0.8047 &	0.2667 & 0.2800 & 0.7984 \\
(Alcaino \citeyear{Alcaino1991}) &  &  &  &  &	 &  &  &  &  &	&  \\
\hline
NGC 4590 & 0.4956 &	0.4500 & 0.4586	& 0.4427 &	0.4693 & 0.7291 & 0.7512 & 0.4513 & 0.1920 &	0.2850 & 0.4422 \\
(Walker \citeyear{Walker1994}) &  &  &  &  &	 &  &  &  &  &	&  \\
\hline
NGC 4833 & 0.6269 &	0.6752 & 0.6533 & 0.6653 & 0.6596 &	0.8189 & 0.8233 & 0.6741 & 0.2131 & 0.2181 & 0.6619 \\
(Melbourne \citeyear{Melbourne2000}) &  &  &  &  &	 &  &  &  &  &	&  \\
\hline
NGC 5466 & 0.4269 &	0.3994 & 0.4036 & 0.3908 & 0.4135 &	0.6733 & 0.6791 & 0.4000 & 0.1939 & 0.2789 & 0.3929 \\
(Buonanno \citeyear{Buonanno1984}) &  &  &  &  &	 &  &  &  &  &	&  \\
\hline
NGC 5694 & 0.5461 &	0.5471 & 0.5382	& 0.5357 & 0.5317 & 0.6657 & 0.6714 & 0.5329 & 0.3624 &	0.4077 & 0.5416 \\
(Ortolani \citeyear{Ortolani1990}) &  &  &  &  &	 &  &  &  &  &	&  \\
\hline
NGC 5927 & 0.9033 &	0.9080 & 0.9021 & 0.9078 & 0.9033 &	0.9365 & 0.9418 & 0.9079 & 0.3487 & 0.4026 & 0.9077 \\
(Samus \citeyear{Samus1996}) &  &  &  &  &	 &  &  &  &  &	&  \\
\hline
NGC 6121 &	0.8348 & 0.8463 & 0.8442 & 0.8465 &	0.8459 & 0.9131 & 0.9147 & 0.8480 & 0.3208 &	0.4010 & 0.8447 \\
(Kanatas \citeyear{Kanatas1995}) &  &  &  &  &	 &  &  &  &  &	&  \\
\hline
NGC 6171 & 0.9268 &	0.9261 & 0.9267 & 0.9242 & 0.9267 &	0.9686 & 0.9708 & 0.9276 & 0.9401 &	0.9314 & 0.9256 \\
(Ferraro \citeyear{Ferraro1991}) &  &  &  &  &	 &  &  &  &  &	&  \\
\hline
NGC 6205 & 0.3276 &	0.3556 & 0.3261 & 0.3458 & 0.3391 &	0.3822 & 0.3739 & 0.3528 &	0.1494 & 0.1801 & 0.3462 \\
(Cohen \citeyear{Cohen1997}) &  &  &  &  &	 &  &  &  &  &	&  \\
\hline
NGC 6218 & 0.5194 &	0.5693 & 0.5337 & 0.5761 & 0.5558 & 0.6059 & 0.6085 & 0.5722 & 0.3230 &	0.3006 & 0.5721 \\
(Sato \citeyear{Sato1989}) &  &  &  &  &	 &  &  &  &  &	&  \\
\hline
NGC 6254 & 0.6713 &	0.7139 & 0.6722 & 0.7142 & 0.6927 &	0.7183 & 0.7335 & 0.7160 & 0.2532 &	0.3165 & 0.7124 \\
(Hurley \citeyear{Hurley1989}) &  &  &  &  &	 &  &  &  &  &	&  \\
\hline
NGC 6352 & 0.8565 &	0.8555 & 0.8536 & 0.8561 & 0.8561 &	0.8974 & 0.9032 & 0.8557 & 0.2981 &	0.2662 & 0.8552 \\
(Fullton \citeyear{Fullton1995}) &  &  &  &  &	 &  &  &  &  &	&  \\
\hline
NGC 6356 & 0.9176 &	0.9208 & 0.9219 & 0.9197 & 0.9229 & 0.9575 & 0.9529 & 0.9208 &	0.3718 & 0.2662 & 0.9193 \\
(Bica \citeyear{Bica1994}) &  &  &  &  &	 &  &  &  &  &	&  \\
\hline
NGC 6362 & 0.7423 & 0.7589 & 0.7496 & 0.7458 &	0.7624 &	0.8285 & 0.7922 & 0.7593 & 0.4152 &	0.4846 & 0.7452 \\
(Brocato \citeyear{Brocato1999}) &  &  &  &  &	 &  &  &  &  &	&  \\
\hline
NGC 6397 &	0.5459 & 0.5494 & 0.5312 & 0.5520 &	0.5548 & 0.6191 & 0.6376 & 0.5512 & 0.2805 &	0.2796 & 0.5487 \\
(Alcaino \citeyear{Alcaino97}) &  &  &  &  &	 &  &  &  &  &	&  \\
\hline
NGC 6553 &	0.9954 &  0.9999 & 0.9940 &	0.9999 & 0.9988 & 0.9879 & 0.9879 & 0.9999 &	0.4247 & 0.4996 & 0.9999 \\
(Sagar \citeyear{Sagar1999}) &  &  &  &  &	 &  &  &  &  &	&  \\
\hline
NGC 6624 & 0.4130 &	0.4090 & 0.4105 & 0.4228 & 0.4305 & 0.5118 & 0.5326 & 0.4147 & 0.3449 &	0.3001 & 0.4162 \\
(Heasley \citeyear{Heasley2000}) &  &  &  &  &	 &  &  &  &  &	&  \\
\hline
NGC 6637 & 0.8578 &	0.8740 & 0.8622 & 0.8737 & 0.8686 & 0.9331 & 0.9374 & 0.8755 &	0.2825 & 0.2554 & 0.8727 \\
(Heasley \citeyear{Heasley2000}) &  &  &  &  &	 &  &  &  &  &	&  \\
\hline
NGC 6712 & 0.8938 & 0.8935 & 0.8963 & 0.8947 & 0.8989 & 0.9632 & 0.9645 & 0.8951 &	0.3044	& 0.3487 &	0.8948 \\
(Ortolani \citeyear{Ortolani2000}) &  &  &  &  &	 &  &  &  &  &	&  \\
\hline
NGC 6838 & 0.7474 &	0.7394 & 0.7393 & 0.7421 & 0.7477 & 0.7626 & 0.7773 & 0.7408 & 0.1744 &	0.1569 & 0.7392 \\
(Hodder \citeyear{Hodder1992}) &  &  &  &  &	 &  &  &  &  &	&  \\
\hline
NGC 7006 & 0.5745 &	0.5595 & 0.5520 & 0.5440 & 0.5568 &	0.6680 & 0.6688 & 0.5518 & 0.3603 &	0.4467 & 0.5455 \\
(Buonanno \citeyear{Buonanno1991}) &  &  &  &  &	 &  &  &  &  &	&  \\
\hline
NGC 7078 & 0.3384 &	0.3615 & 0.3397 & 0.3539 & 0.3507 &	0.5045 & 0.4809 & 0.3630 & 0.2426 &	0.1924 & 0.3528 \\
(Stetson \citeyear{Stetson1994}) &  &  &  &  &	 &  &  &  &  &	&  \\
\hline
NGC 7099  &	0.5447 & 0.5127 & 0.5132 & 0.5042 &	0.5273 &	0.6886 & 0.6825 & 0.5183 & 0.3048 &	0.3116 & 0.5063 \\
(Buonanno \citeyear{Buonanno1988}) &  &  &  &  &	 &  &  &  &  &	&  \\
\hline
NGC 7492 & 0.5509 &	0.5338 & 0.5296 & 0.5212 & 0.5237 &	0.5975 & 0.5927 & 0.5218 & 0.3506 &	0.4406 & 0.5238 \\
(Cote \citeyear{Cote1991}) &  &  &  &  &	 &  &  &  &  &	&  \\
\hline
NGC 2419 & 0.4649 &	0.4902 & 0.4695 & 0.4805 & 0.4878 &	0.6296 & 0.6186 & 0.4920 & 0.2037 &	0.2132 & 0.4796 \\
(Harris \citeyear{Harris1997}) &  &  &  &  &	 &  &  &  &  &	&  \\
\hline
NGC 5272 & 0.3743 & 0.3747 & 0.3508 & 0.3632 & 0.3646 & 0.4882 & 0.5160 & 0.3702 & 0.2120 &	0.2466 & 0.3664 \\
(Buonanno \citeyear{Buonanno1994}) &  &  &  &  &	 &  &  &  &  &	&  \\
\hline
$\omega$ Centauri & 0.4593 & 0.4816 & 0.4596 & 0.4764 & 0.4792 & 0.6099 & 0.6235 & 0.4893 & 0.1886 & 0.2097 & 0.4769 \\
(Castellani \citeyear{Castellani2007}) &  &  &  &  &	 &  &  &  &  &	&  \\
\hline
Palomar 4 & 0.5647 & 0.5542 & 0.5496 &	0.5399 & 0.5516 & 0.7566 & 0.7737 & 0.5449 & 0.4257 & 0.3374 & 0.5395 \\
(Christian \citeyear{Christian1986}) &  &  &  &  &	 &  &  &  &  &	&  \\
 \hline
\end{tabular}}
\end{table}

\begin{table}
\renewcommand{\arraystretch}{0.8}
\centering
\caption{ Same as Table \ref{tab:2} but for observed SMC  star clusters.}\label{tab:3}
\begin{tabular}{c|c|c|c|c|c|c|c|c|c|c|c}
\hline
Names & DPL & DPL & $\beta$ & $\beta$ & DE & E & E & G3 & G3 & G2 & G2 \\
of & ($\alpha$, $\beta$) & ($\alpha$, $\beta$) & (p, q)  & (p, q)  & ($T_{0}$, $\tau$) & ($\lambda$) & ($\lambda$) & ($\mu_{1}$, $\sigma_{1}$) & ($\mu_{1}$, $\sigma_{1}$) &  ($\mu_{1}$, $\sigma_{1}$) & ($\mu_{1}$, $\sigma_{1}$) \\ 
observed & $(\tau$) & ($\tau$) &  &  &  &  &  & ($\mu_{2}$, $\sigma_{2}$)  & ($\mu_{2}$, $\sigma_{2}$) & ($\mu_{2}$, $\sigma_{2}$) & ($\mu_{2}$, $\sigma_{2}$) \\
star & (50, 0.2) & (50, 10)  & (2, 2) & (5, 1) & (15, 9) & (5) & (10) & ($\mu_{3}$, $\sigma_{3}$) & ($\mu_{3}$, $\sigma_{3}$) & (q) & (q) \\
clusters & (15) & (15) &  &  &  &  &  & ($q_{1}$, $q_{2}$) & ($q_{1}$, $q_{2}$)  &  &  \\
 &  &   &  &  &  &  &  & (9, 0.05) & (2, 0.05) & (2, 0.2) &  (11, 0.2) \\
 &  &  &  &  &  &  &  & (11, 0.05) & (5, 0.05) & (8, 0.5) & (13, 0.5) \\
  &  &  &  &  &  &  &  & (13, 0.05) & (7, 0.05) & (0.5) & (0.5)\\
&  &  &  &  &  &  &  & (0.5, 0.3) & (0.5, 0.3) &  &  \\
\hline
SMC Br\"{u}ck 2 & 0.4845 & 0.5473 & 0.4920 & 0.5462 & 0.5102 & 0.5643 & 0.5571 & 0.5472 & 0.2654 & 0.3364 & 0.5467 \\
(Dias \citeyear{Dias2016}) &  &  &  &  &  &  &  &  &  &  &  \\
\hline
SMC Br\"{u}ck 4 & 0.4603 & 0.5133 &	0.4641 & 0.5196 & 0.4901 & 0.5537 & 0.5534 & 0.5154 & 0.3081 & 0.3620 & 0.5185 \\
(Dias \citeyear{Dias2016}) &  &  &  &  &  &  &  &  &  &  &  \\
\hline
SMC Br\"{u}ck 6 & 0.5034 & 0.5123 &	0.4919 & 0.5141	& 0.5090 & 0.6691 & 0.6885 & 0.5140 & 0.3142 &	0.3659 & 0.5140 \\
(Dias \citeyear{Dias2016}) &  &  &  &  &  &  &  &  &  &  &  \\
\hline
SMC HW 5 & 0.4476 &	0.4759 & 0.4453 & 0.4830 & 0.4720 & 0.5965 & 0.6231 & 0.4778 & 0.3601 & 0.3874 & 0.4802 \\
(Dias \citeyear{Dias2016}) &  &  &  &  &  &  &  &  &  &  &  \\
\hline
SMC HW 6 & 0.4331 & 0.4777 & 0.4398 & 0.4833 & 0.4642 &	0.5960 & 0.5946 & 0.4787 & 0.3259 & 0.3614 & 0.4827 \\
(Dias \citeyear{Dias2016}) &  &  &  &  &  &  &  &  &  &  &  \\
\hline
SMC Kron 11 & 0.4200 & 0.4605 & 0.4166 & 0.4687 & 0.4469 & 0.5059 & 0.5323 & 0.4630 & 0.3368 &	0.3925 & 0.4666 \\
(Dias \citeyear{Dias2016}) &  &  &  &  &  &  &  &  &  &  &  \\
\hline
SMC Kron 8 & 0.4190 & 0.4523 & 0.4045 &	0.4461 & 0.4242 & 0.5814 & 0.6075 & 0.4532 & 0.1850 & 0.2471 & 0.4479 \\
(Dias \citeyear{Dias2016}) &  &  &  &  &  &  &  &  &  &  &  \\
\hline
SMC Lindsay 14 & 0.4420 & 0.4853 & 0.4556 &	0.4916 &	0.4797 & 0.6384 & 0.6376 & 0.4863 &	0.3572 & 0.3814 & 0.4900 \\
(Dias \citeyear{Dias2016}) &  &  &  &  &  &  &  &  &  &  &  \\
\hline
SMC NGC 152 & 0.4727 & 0.4787 &	0.4400 & 0.4762 & 0.4604 & 0.6303 & 0.6603 & 0.4814 & 0.2198 & 0.2863 & 0.4766 \\
(Dias \citeyear{Dias2016}) &  &  &  &  &  &  &  &  &  &  &  \\
\hline
   
\end{tabular}
\end{table}

\begin{table}
\centering
\caption{ Same as Table \ref{tab:2} but for observed MW  open clusters.}\label{tab:4}
\begin{tabular}{c|c|c|c|c|c|c|c|c|c|c|c}
\hline
Names & DPL & DPL & $\beta$ & $\beta$ & DE & E & E & G3 & G3 & G2 & G2 \\
of & ($\alpha$, $\beta$) & ($\alpha$, $\beta$) & (p, q)  & (p, q)  & ($T_{0}$, $\tau$) & ($\lambda$) & ($\lambda$) & ($\mu_{1}$, $\sigma_{1}$) & ($\mu_{1}$, $\sigma_{1}$) &  ($\mu_{1}$, $\sigma_{1}$) & ($\mu_{1}$, $\sigma_{1}$) \\ 
observed & $(\tau$) & ($\tau$) &  &  &  &  &  & ($\mu_{2}$, $\sigma_{2}$)  & ($\mu_{2}$, $\sigma_{2}$) & ($\mu_{2}$, $\sigma_{2}$) & ($\mu_{2}$, $\sigma_{2}$) \\
star & (50, 0.2) & (50, 10)  & (2, 2) & (5, 1) & (15, 9) & (5) & (10) & ($\mu_{3}$, $\sigma_{3}$) & ($\mu_{3}$, $\sigma_{3}$) & (q) & (q) \\
clusters & (15) & (15) &  &  &  &  &  & ($q_{1}$, $q_{2}$) & ($q_{1}$, $q_{2}$)  &  &  \\
 &  &   &  &  &  &  &  & (9, 0.05) & (2, 0.05) & (2, 0.2) &  (11, 0.2) \\
 &  &  &  &  &  &  &  & (11, 0.05) & (5, 0.05) & (8, 0.5) & (13, 0.5) \\
  &  &  &  &  &  &  &  & (13, 0.05) & (7, 0.05) & (0.5) & (0.5)\\
&  &  &  &  &  &  &  & (0.5, 0.3) & (0.5, 0.3) &  &  \\
\hline
H and $\chi$ Persei & 0.8124 & 0.8084 & 0.8111 & 0.8076 & 0.8121 & 0.8271 & 0.8278 &	0.8131 & 0.5978 & 0.5773 & 0.8091 \\
(Waelkens \citeyear{Waelkens1990}) &  &  &  &  &  &  & & &  &  & \\
\hline
NGC 2264 & 0.9093 & 0.9048 & 0.9057 & 0.9038 & 0.9055 & 0.9076 & 0.9070 & 0.9086 & 0.7975 & 0.7908 & 0.9039 \\
(Turner \citeyear{Turner2012}) &  &  &  &  &  &  & & &  &  & \\
\hline
NGC 2547 & 0.7092 & 0.7306 & 0.7214 & 0.7309 & 0.7320 & 0.7020 & 0.6959 & 0.7326 & 0.4539 & 0.4918 & 0.7308 \\
(Naylor \citeyear{Naylor2002}) &  &  &  &  &  &  & & &  &  & \\
\hline
NGC 6811 &	0.5858 & 0.5588 & 0.5686 & 0.5610 & 0.5717 & 0.8654 & 0.8546 & 0.5663 & 0.2631 & 0.3095 & 0.5592 \\
(Yontan \citeyear{Yontan2015}) &  &  &  &  &  &  & & &  &  & \\
\hline
NGC 7160 & 0.9999 & 0.9999 & 0.9999 & 0.9999 & 0.9999 & 0.9999 & 0.9999 & 0.9999 & 0.9998 & 0.9998 & 0.9999\\ 
(Mayne \citeyear{Mayne2007}) &  &  &  &  &  &  & & &  &  & \\
\hline
$\sigma$ Orionis &	0.9988 & 0.9988 & 0.9988 & 0.9988 &	0.9988 & 0.9988 & 0.9988 & 0.9988 & 0.9988 & 0.9988 & 0.9988 \\
(Bejar \citeyear{Bejar2004}) &  &  &  &  &  &  & & &  &  & \\
\hline

\end{tabular}
\end{table}

\begin{table}
\renewcommand{\arraystretch}{0.95}
\centering
\caption{Same as Table \ref{tab:2} but for various LFs other than Gaia}\label{tab:5}

\begin{tabular}{c|c|c|c|c|c}
\hline
Names & G2 & E & E &  DPL & G2 \\
of & ($\mu_{1}$, $\sigma_{1}$) & ($\lambda$) & ($\lambda$) & ($\alpha$, $\beta$)  & ($\mu_{1}$, $\sigma_{1}$) \\
observed & ($\mu_{2}$, $\sigma_{2}$) &  (1) & (1) & ($\tau$) & ($\mu_{2}$, $\sigma_{2}$) \\
star &  (q) &  &  & (50, 10) & (q) \\
clusters & (3, 0.5) &  &  & (15) & (11, 0.2) \\
 & (8, 0.9) &  &  &  & (13, 0.5) \\
 & (0.6) &  &  &  & (0.5) \\
 & LF:=exp & LF:=exp & LF:=exp &  LF:= \footnote{TPL stands for Truncated Power Law luminosity function}TPL  & LF:=TPL \\
& ($\lambda$=10) & ($\lambda$=2.05) & ($\lambda$=10) & ($\alpha$=1.05) &  ($\alpha$=1.05) \\
&  &  &  & ($\beta$=8) & ($\beta$=8) \\
&  &  &  & ($\nu$=2) & ($\nu$=4) \\
\hline
47 Tucanae & 0.4255 & 0.8406 & 0.6237 & 0.4659 &	0.6042 \\
\hline
M 14 & 0.8345 &	0.9593 & 0.8879 & 0.7792 & 0.8569 \\
\hline
M 5 & 0.3055 & 0.7460 &	0.6070 & 0.3261 & 0.4946 \\
\hline
M 92 & 0.6574 &	0.8653 & 0.8094 & 0.3430 & 0.6519 \\
\hline
M 53 & 0.7757 &	0.9292 & 0.8639 & 0.8394 & 0.8859 \\
\hline
NGC 1261 & 0.8615 &	0.9624 & 0.9561 & 0.6409 & 0.8418 \\
\hline
NGC 1904 & 0.9973 &	0.9975 & 0.8949 & 0.8790 & 0.9973 \\
\hline
NGC 2808 & 0.8032 &	0.9516 & 0.8887 & 0.6731 & 0.8038 \\
\hline
NGC 288 & 0.8262 & 0.9577 &	0.8839 & 0.6338 & 0.8237 \\
\hline
NGC 3201 & 0.8483 &	0.9722 & 0.8699 & 0.7102 & 0.8611 \\
\hline
NGC 362 & 0.8604 & 0.9777 &	0.9144 & 0.6920 & 0.8446 \\
\hline
NGC 4147 & 0.8272 &	0.9705 & 0.9076 & 0.5577 & 0.7862 \\
\hline
NGC 4372 & 0.8811 &	0.9763 & 0.9025 & 0.7989 & 0.8832 \\
\hline
NGC 4590 & 0.7391 &	0.9556 & 0.8900 & 0.4320 & 0.7294 \\
\hline
NGC 4833 & 0.7515 &	0.9466 & 0.8838 & 0.6530 & 0.8097 \\
\hline
NGC 5466 & 0.7672 &	0.9597 & 0.8896 & 0.3992 & 0.7225 \\
\hline
NGC 5694 & 0.8641 &	0.9764 & 0.9701 & 0.5403 & 0.8186 \\
\hline
NGC 5927 & 0.9244 &	0.9835 & 0.9169 & 0.9085 & 0.9237 \\
\hline
NGC 6121 & 0.8864 &	0.9697 & 0.9153 & 0.8346 & 0.8920 \\
\hline
NGC 6171 & 0.9433 & 0.9937 & 0.9931 & 0.9240 & 0.9455 \\
\hline
NGC 6205 & 0.4856 &	0.7509 & 0.6827 & 0.3634 & 0.5757 \\
\hline
NGC 6218 & 0.5462 &	0.8611 & 0.7625 & 0.5590 & 0.6991 \\
\hline
NGC 6254 & 0.7059 &	0.8822 & 0.8610 & 0.7169 & 0.7876 \\
\hline
NGC 6352 & 0.8818 &	0.9621 & 0.8438 & 0.8579 & 0.8823 \\
\hline
NGC 6356 & 0.9433 &	0.9932 & 0.9254 & 0.9163 & 0.9419 \\
\hline
NGC 6362 & 0.9022 &	0.9781 & 0.9667 & 0.7054 & 0.9012 \\
\hline
NGC 6397 & 0.6015 &	0.8817 & 0.8206 & 0.5633 & 0.6958 \\
\hline
NGC 6553 & 0.9930 &	0.9936 & 0.9116 & 0.9999 & 1.0000 \\
\hline
NGC 6624 & 0.3267 &	0.7916 & 0.5926 & 0.4233 & 0.5556 \\
\hline
NGC 6637 & 0.9011 &	0.9809 & 0.9073 & 0.8673 & 0.9123 \\
\hline
NGC 6712 & 0.9238 &	0.9906 & 0.9220 & 0.8866 & 0.9234 \\
\hline
NGC 6838 & 0.7035 &	0.8789 & 0.6391 & 0.7443 & 0.7830 \\
\hline
NGC 7006 & 0.8614 &	0.9681 & 0.9640 & 0.5780 & 0.8253 \\
\hline
NGC 7078 & 0.6575 &	0.8803 & 0.8135 & 0.3569 & 0.6577 \\
\hline
NGC 7099 & 0.7888 &	0.9709 & 0.8752 & 0.5117 & 0.7779 \\
\hline
NGC 7492 & 0.8627 &	0.9681 & 0.9633 & 0.5641 & 0.8012 \\
\hline
NGC 2419 & 0.7620 &	0.9459 & 0.8691 & 0.4852 & 0.7509 \\
\hline
NGC 5272 & 0.6296 &	0.8612 & 0.8134 & 0.4025 & 0.6563 \\
\hline
$\omega$ Centauri & 0.7078 & 0.9046 & 0.8245 &	0.4547 & 0.7179 \\
\hline
Palomar 4 &	0.8412 & 0.9814 & 0.9255 & 0.5542 &	0.8142 \\

\end{tabular}
\end{table}

\begin{table}
\centering
\caption{Same as Table \ref{tab:5} but for observed SMC star clusters.}\label{tab:6}

\begin{tabular}{c|c|c|c|c|c}
\hline
Names & G2 & E & E &  DPL & G2 \\
of & ($\mu_{1}$, $\sigma_{1}$) & ($\lambda$) & ($\lambda$) & ($\alpha$, $\beta$)  & ($\mu_{1}$, $\sigma_{1}$) \\
observed & ($\mu_{2}$, $\sigma_{2}$) &  (1) & (1) & ($\tau$) & ($\mu_{2}$, $\sigma_{2}$) \\
star &  (q) &  &  & (50, 10) & (q) \\
clusters & (3, 0.5) &  &  & (15) & (11, 0.2) \\
 & (8, 0.9) &  &  &  & (13, 0.5) \\
 & (0.6) &  &  &  & (0.5) \\
 & LF:=exp & LF:=exp & LF:=exp &  LF:=TPL  & LF:=TPL \\
& ($\lambda$=10) & ($\lambda$=2.05) & ($\lambda$=10) & ($\alpha$=1.05) &  ($\alpha$=1.05) \\
&  &  &  & ($\beta$=8) & ($\beta$=8) \\
&  &  &  & ($\nu$=2) & ($\nu$=4) \\
\hline
SMC Br\"{u}ck 2 & 0.6407 & 0.8942 &	0.8578 & 0.5272 & 0.7427 \\
\hline
SMC Br\"{u}ck 4 & 0.5771 & 0.8801 &	0.8601 & 0.4990 & 0.7054 \\
\hline
SMC Br\"{u}ck 6 & 0.6556 & 0.9053 &	0.8636 & 0.5039 & 0.7100 \\
\hline
SMC HW 5 & 0.5318 &	0.8662 & 0.8655 & 0.4727 & 0.6663 \\
\hline
SMC HW 6 & 0.5806 &	0.8873 & 0.8657 & 0.4537 & 0.6916 \\
\hline
SMC Kron 11 & 0.5161 & 0.8512 &	0.8574 & 0.4589 & 0.6611 \\
\hline
SMC Kron 8 & 0.6576 & 0.9173 & 0.8582 &	0.4508 & 0.7207 \\
\hline
SMC Lindsay 14 & 0.5831 & 0.8805 & 0.8690 &	0.4600 & 0.6884 \\
\hline
SMC NGC 152 & 0.6807 & 0.9182 &	0.8599 & 0.4861 & 0.7279  \\
\hline

\end{tabular}
\end{table}

\begin{table}
\centering
\caption{Same as Table \ref{tab:5} but for observed MW  open clusters.}\label{tab:7}
    
\begin{tabular}{c|c|c|c|c|c}
\hline
Names & G2 & E & E &  DPL & G2 \\
of & ($\mu_{1}$, $\sigma_{1}$) & ($\lambda$) & ($\lambda$) & ($\alpha$, $\beta$)  & ($\mu_{1}$, $\sigma_{1}$) \\
observed & ($\mu_{2}$, $\sigma_{2}$) &  (1) & (1) & ($\tau$) & ($\mu_{2}$, $\sigma_{2}$) \\
star &  (q) &  &  & (50, 10) & (q) \\
clusters & (3, 0.5) &  &  & (15) & (11, 0.2) \\
 & (8, 0.9) &  &  &  & (13, 0.5) \\
 & (0.6) &  &  &  & (0.5) \\
 & LF:=exp & LF:=exp & LF:=exp &  LF:=TPL  & LF:=TPL \\
& ($\lambda$=10) & ($\lambda$=2.05) & ($\lambda$=10) & ($\alpha$=1.05) &  ($\alpha$=1.05) \\
&  &  &  & ($\beta$=8) & ($\beta$=8) \\
&  &  &  & ($\nu$=2) & ($\nu$=4) \\
\hline
H and $\chi$ Persei & 0.8342 & 0.6693 &	0.2524 & 0.8090 & 0.6678 \\
\hline
NGC 2264 & 0.9478 &	0.7608 & 0.4425 & 0.9043 & 0.8255 \\
\hline
NGC 2547 & 0.7307 &	0.7678 & 0.1736 & 0.7190 & 0.7421 \\
\hline
NGC 6811 & 0.7725 &	0.9644 & 0.9008 & 0.5211 & 0.7795 \\
\hline
NGC 7160 & 0.9999 &	0.9999 & 0.9998 & 0.9999 & 0.9999 \\
\hline
$\sigma$ Orionis &	0.9988 & 0.9988 & 0.9988 & 0.9988 &	0.9988 \\
\hline
  
\end{tabular}
\end{table}

\begin{table}
\centering
\renewcommand{\arraystretch}{0.95}
\caption{Same as Table \ref{tab:2} but for Z=0.019 }\label{tab:8}

\begin{tabular}{c|c|c|c|c|c|c|c|c|c|c}
\hline
Names & $\beta$ & $\beta$ & DE & DPL & E & E & G2 & G2 & G3 & G3 \\
of & (p, q) &  (p, q)  & ($T_{0}$, $\tau$) & ($\alpha$, $\beta$) & ($\lambda$) & ($\lambda$)  & ($\mu_{1}$, $\sigma_{1}$)  & ($\mu_{1}$, $\sigma_{1}$)  & ($\mu_{1}$, $\sigma_{1}$)  & ($\mu_{1}$, $\sigma_{1}$)  \\
observed & (2, 2) & (5, 1) & (15, 2) & ($\tau$) & (5) & (10) & ($\mu_{2}$, $\sigma_{2}$) & ($\mu_{2}$, $\sigma_{2}$) & ($\mu_{2}$, $\sigma_{2}$) & ($\mu_{2}$, $\sigma_{2}$)  \\
star  &  &  &  & (50, 10) &  &  & (q) & (q) & ($\mu_{3}$, $\sigma_{3}$) & ($\mu_{3}$, $\sigma_{3}$) \\
clusters  &  &  &  & (15) &  &  & (11, 0.2) & (3, 0.5) & ($q_{1}$, $q_{2}$) & ($q_{1}$, $q_{2}$) \\
 &  &  &  &  &  &  & (13, 0.5) & (8, 0.9)  & (2, 0.05) & (9, 0.05) \\
 &  &  &  &  &  &  & (0.5) & (0.6)  & (5, 0.05) & (11, 0.05) \\
 & &  &  &  & &  &   &  & (7, 0.05) & (13, 0.05) \\
 &  &  &  & &  &  &   &  & (0.5, 0.3) & (0.5, 0.3) \\
\hline
47 Tucanae & 0.5126 & 0.1620 & 0.4100 & 0.0843 &	0.3307 & 0.4022 & 0.4763 & 0.3709 & 0.0859 & 0.4598 \\
\hline
M 14 & 0.7801 &	0.3770 & 0.6424 & 0.2644 & 0.6023 &	0.5380 &	0.6432 & 0.6272 & 0.2709 & 0.6543 \\
\hline
M 5 & 0.3564 & 0.0779 &	0.4806 & 0.0986 & 0.2318 & 0.2984 &	0.5597 & 0.4303 & 0.1039 & 0.5442 \\
\hline
M 92 & 0.4771 &	0.1839 & 0.8225 & 0.3094 & 0.3212 &	0.4157 &	0.8777 & 0.7917 & 0.3049 & 0.8756 \\
\hline
M 53 & 0.8281 &	0.5140 & 0.8417 & 0.6404 & 0.4334 &	0.5651 &	0.8782 & 0.8147 & 0.6378 & 0.8805 \\
\hline
NGC 1261 & 0.6182 &	0.2748 & 0.7371 & 0.4133 & 0.5095 &	0.7399 & 0.7439 & 0.7133 & 0.4150 &	0.7450 \\
\hline
NGC 1904 & 0.8790 &	0.8813 & 0.8789 & 0.8784 & 0.8813 &	0.8784 & 0.8788 & 0.8789 & 0.8789 &	0.8788 \\
\hline
NGC 2808 & 0.6911 &	0.2153 & 0.3864 & 0.2272 & 0.5665 &	0.6305 & 0.4122 & 0.3936 & 0.2267 &	0.4141 \\
\hline
NGC 288 & 0.6268 & 0.2106 &	0.5842 & 0.3598 & 0.4686 &	0.6799 & 0.6257 & 0.5748 & 0.3455 &	0.6119 \\
\hline
NGC 3201 & 0.7324 &	0.3359 & 0.6162 & 0.2248 & 0.5378 &	0.4819 & 0.6338 & 0.6139 & 0.2327 &	0.6201 \\
\hline
NGC 362 & 0.7627 & 0.2685 &	0.5837 & 0.3259 & 0.6040 &	0.8262 & 0.6217 & 0.5761 & 0.3183 &	0.6043 \\
\hline
NGC 4147 & 0.6206 &	0.2504 & 0.8275 & 0.3583 & 0.3105 &	0.5175 & 0.8672 & 0.8127 & 0.3464 &	0.8507 \\
\hline
NGC 4372 & 0.8070 &	0.3266 & 0.4356 & 0.2012 & 0.6177 &	0.6103 & 0.4422 & 0.4675 & 0.1992 &	0.4459 \\
\hline
NGC 4590 & 0.5459 &	0.2598 & 0.7749 & 0.1909 & 0.5330 &	0.4402 & 0.8115 & 0.7147 & 0.1963 & 0.8222 \\
\hline
NGC 4833 & 0.6777 &	0.2677 & 0.4346 & 0.2118 & 0.5440 &	0.5839 & 0.4778 & 0.4216 & 0.2147 &	0.4663 \\
\hline
NGC 5466 & 0.6407 &	0.2310 & 0.8922 & 0.2158 & 0.4651 &	0.3997 & 0.9396 & 0.8689 & 0.2125 &	0.9371 \\
\hline
NGC 5694 & 0.7365 &	0.4251 & 0.8466 & 0.3708 & 0.7077 &	0.8400 & 0.8691 & 0.8411 & 0.3792 &	0.8574 \\
\hline
NGC 5927 & 0.9157 &	0.4319 & 0.6490 & 0.2897 & 0.6738 &	0.5664 & 0.6364 & 0.7018 & 0.2941 &	0.6327 \\
\hline
NGC 6121 & 0.8779 &	0.3420 & 0.5260 & 0.2535 & 0.5401 &	0.5853 & 0.4916 & 0.5976 & 0.2780 &	0.4953 \\
\hline
NGC 6171 & 0.9174 &	0.8452 & 0.6061 & 0.6128 & 0.9682 &	0.9771 & 0.5955 & 0.6269 & 0.6484 &	0.6111 \\
\hline
NGC 6205 & 0.3773 &	0.1044 & 0.6148 & 0.2240 & 0.2398 &	0.3337 & 0.6678 & 0.5874 & 0.2316 &	0.6643 \\
\hline
NGC 6218 & 0.4872 &	0.1866 & 0.4462 & 0.3412 & 0.3035 &	0.4647 & 0.4884 & 0.4228 & 0.3344 &	0.4743 \\
\hline
NGC 6254 & 0.6542 &	0.2002 & 0.4053 & 0.2884 & 0.3501 &	0.4976 & 0.3843 & 0.4303 & 0.2881 &	0.3992 \\
\hline
NGC 6352 & 0.8749 &	0.2207 & 0.5261 & 0.2080 & 0.4442 &	0.5958 & 0.4917 & 0.5850 & 0.2087 &	0.5120 \\
\hline
NGC 6356 & 0.8973 &	0.2558 & 0.5597 & 0.2477 & 0.6340 &	0.7761 & 0.5342 & 0.5919 & 0.2280 &	0.5532 \\
\hline
NGC 6362 & 0.6960 &	0.2129 & 0.6099 & 0.4539 & 0.5291 &	0.7990 & 0.6202 & 0.5868 & 0.4638 &	0.6346 \\
\hline
NGC 6397 & 0.4521 &	0.1842 & 0.4183 & 0.3049 & 0.3031 &	0.3989 & 0.4729 & 0.4052 & 0.3081 &	0.4547 \\
\hline
NGC 6553 & 0.9785 &	0.4583 & 0.9163 & 0.3666 & 0.6511 &	0.5553 & 0.8988 & 0.9289 & 0.3912 &	0.9026 \\
\hline
NGC 6624 & 0.5012 &	0.2538 & 0.6876 & 0.2580 & 0.3464 &	0.3361 & 0.7674 & 0.6155 & 0.2604 &	0.7441 \\
\hline
NGC 6637 & 0.8685 &	0.2917 & 0.5516 & 0.2258 & 0.6191 & 0.6811 & 0.5539 & 0.5719 & 0.2058 &	0.5559 \\
\hline
NGC 6712 & 0.8917 &	0.4172 & 0.6797 & 0.2346 & 0.6681 &	0.6142 & 0.6358 & 0.6536 & 0.2348 & 0.6783 \\
\hline
NGC 6838 & 0.6876 &	0.1449 & 0.3837 & 0.0672 & 0.3371 &	0.4474 & 0.4030 & 0.4332 & 0.0723 &	0.3923 \\
\hline
NGC 7006 & 0.6480 &	0.2316 & 0.7352 & 0.4154 & 0.5027 &	0.7430 & 0.7686 & 0.7248 & 0.4179 &	0.7574 \\
\hline
NGC 7078 & 0.4893 &	0.1672 & 0.7861 & 0.2372 & 0.3466 &	0.4392 & 0.8269 & 0.7552 & 0.2407 &	0.8270 \\
\hline
NGC 7099 & 0.6586 &	0.2889 & 0.9099 & 0.3049 & 0.4176 &	0.4105 & 0.9448 & 0.8896 & 0.3055 & 0.9355 \\
\hline
NGC 7492 & 0.6066 &	0.4304 & 0.8018 & 0.4131 & 0.4788 &	0.6178 & 0.8381 & 0.7798 & 0.4137 & 0.8284 \\
\hline
NGC 2419 & 0.5208 &	0.2020 & 0.6686 & 0.2257 & 0.4352 &	0.4510 & 0.7079 & 0.6377 & 0.2189 & 0.7004 \\
\hline
NGC 5272 & 0.4364 &	0.1584 & 0.6583 & 0.2546 & 0.3241 &	0.4213 & 0.7199 & 0.6125 & 0.2533 & 0.7163 \\
\hline
$\omega$ Centauri & 0.5277 & 0.2742 & 0.7885 &	0.2462 &	0.4021 & 0.3264 & 0.8331 & 0.7558 &	0.2445 & 0.8303 \\
\hline
Palomar 4 &	0.7148 & 0.2874 & 0.6517 & 0.4201 &	0.6430 &	0.8754 & 0.6777 & 0.6305 & 0.4190 &	0.6726 \\
     
\end{tabular}
\end{table}

\begin{table}
\centering
\caption{Same as Table \ref{tab:8} but for observed SMC star clusters}\label{tab:9}
    
\begin{tabular}{c|c|c|c|c|c|c|c|c|c|c}
\hline
Names & $\beta$ & $\beta$ & DE & DPL & E & E & G2 & G2 & G3 & G3 \\
of & (p, q) &  (p, q)  & ($T_{0}$, $\tau$) & ($\alpha$, $\beta$) & ($\lambda$) & ($\lambda$)  & ($\mu_{1}$, $\sigma_{1}$)  & ($\mu_{1}$, $\sigma_{1}$)  & ($\mu_{1}$, $\sigma_{1}$)  & ($\mu_{1}$, $\sigma_{1}$)  \\
observed & (2, 2) & (5, 1) & (15, 2) & ($\tau$) & (5) & (10) & ($\mu_{2}$, $\sigma_{2}$) & ($\mu_{2}$, $\sigma_{2}$) & ($\mu_{2}$, $\sigma_{2}$) & ($\mu_{2}$, $\sigma_{2}$)  \\
star  &  &  &  & (50, 10) &  &  & (q) & (q) & ($\mu_{3}$, $\sigma_{3}$) & ($\mu_{3}$, $\sigma_{3}$) \\
clusters  &  &  &  & (15) &  &  & (11, 0.2) & (3, 0.5) & ($q_{1}$, $q_{2}$) & ($q_{1}$, $q_{2}$) \\
 &  &  &  &  &  &  & (13, 0.5) & (8, 0.9)  & (2, 0.05) & (9, 0.05) \\
 &  &  &  &  &  &  & (0.5) & (0.6)  & (5, 0.05) & (11, 0.05) \\
 & &  &  &  & &  &   &  & (7, 0.05) & (13, 0.05) \\
 &  &  &  & &  &  &   &  & (0.5, 0.3) & (0.5, 0.3) \\
\hline
SMC Br\"{u}ck 2 &	0.5815 &	0.3898 &	0.8488 &	0.4071 &	0.4143 &	0.4665 &	0.9098 &	0.7849 &	0.4078 &	0.9063 \\
\hline
SMC Br\"{u}ck 4 &	0.5503 &	0.3200 &	0.8028 &	0.4060 &	0.3459 &	0.4499 &	0.8866 &	0.7287 &	0.4063 &	0.8619 \\
\hline
SMC Br\"{u}ck 6 &	0.5113 &	0.3462 &	0.6528 &	0.3436 &	0.5041 &	0.4326 &	0.6906 &	0.6179 &	0.3438 &	0.6876 \\
\hline
SMC HW 5 &	0.5266 &	0.2161 &	0.7557 &	0.4028 &	0.3242 &	0.4698 &	0.8163 &	0.6648 &	0.4018 &	0.8137 \\
\hline
SMC HW 6 &	0.5334 &	0.2628 &	0.7939 &	0.3776 &	0.3704 &	0.4588 &	0.8868 &	0.7177 &	0.3751 & 	0.8553 \\
\hline
SMC Kron 11 &	0.4803 &	0.2167 &	0.7937 &	0.4200 &	0.2857 &	0.4306 &	0.8521 &	0.7018 &	0.4200 &	0.8508 \\
\hline
SMC Kron 8 &	0.5985 &	0.2941 &	0.8947 &	0.2988 &	0.4203 &	0.4775 &	0.9373 &	0.8344 &	0.2986 &	0.9335 \\
\hline
SMC Lindsay 14 &	0.5194 &	0.2385 &	0.7638 &	0.3795 &	0.3786 &	0.4754 &	0.8487 &	0.6758 &	0.3788 &	0.8257 \\
\hline
SMC NGC 152 &	0.5316 &	0.3242 &	0.8275 &	0.3014 &	0.4792 &	0.4128 &	0.8990 &	0.7791 &	0.3025 &	0.8735 \\
\hline
     
\end{tabular}
\end{table}

\begin{table}
\centering
\caption{ Same as Table \ref{tab:8} but for observed MW open clusters}\label{tab:10}
    
\begin{tabular}{c|c|c|c|c|c|c|c|c|c|c}
\hline
Names & $\beta$ & $\beta$ & DE & DPL & E & E & G2 & G2 & G3 & G3 \\
of & (p, q) &  (p, q)  & ($T_{0}$, $\tau$) & ($\alpha$, $\beta$) & ($\lambda$) & ($\lambda$)  & ($\mu_{1}$, $\sigma_{1}$)  & ($\mu_{1}$, $\sigma_{1}$)  & ($\mu_{1}$, $\sigma_{1}$)  & ($\mu_{1}$, $\sigma_{1}$)  \\
observed & (2, 2) & (5, 1) & (15, 2) & ($\tau$) & (5) & (10) & ($\mu_{2}$, $\sigma_{2}$) & ($\mu_{2}$, $\sigma_{2}$) & ($\mu_{2}$, $\sigma_{2}$) & ($\mu_{2}$, $\sigma_{2}$)  \\
star  &  &  &  & (50, 10) &  &  & (q) & (q) & ($\mu_{3}$, $\sigma_{3}$) & ($\mu_{3}$, $\sigma_{3}$) \\
clusters  &  &  &  & (15) &  &  & (11, 0.2) & (3, 0.5) & ($q_{1}$, $q_{2}$) & ($q_{1}$, $q_{2}$) \\
 &  &  &  &  &  &  & (13, 0.5) & (8, 0.9)  & (2, 0.05) & (9, 0.05) \\
 &  &  &  &  &  &  & (0.5) & (0.6)  & (5, 0.05) & (11, 0.05) \\
 & &  &  &  & &  &   &  & (7, 0.05) & (13, 0.05) \\
 &  &  &  & &  &  &   &  & (0.5, 0.3) & (0.5, 0.3) \\
\hline
H and $\chi$ Persei & 0.7370 & 0.5813 &	0.7362 & 0.5705 &	0.5866 & 0.5927 & 0.7375 & 0.7309 &	0.5759 & 0.7514 \\
\hline
NGC 2264 & 0.8783 &	0.7940 & 0.8771 & 0.7916 & 0.7892 &	0.7925 & 0.8773 & 0.8755 & 0.7927 &	0.8812 \\
\hline
NGC 2547 & 0.6111 &	0.4524 & 0.7348 & 0.4726 & 0.4062 &	0.4305 & 0.7695 & 0.7058 & 0.4770 &	0.7634 \\
\hline
NGC 6811 & 0.6605 &	0.4131 & 0.6902 & 0.2247 & 0.6428 &	0.5624 & 0.7299 & 0.6419 & 0.2373 &	0.7287 \\
\hline
NGC 7160 & 0.9999 &	0.9998 & 0.9998 & 0.9998 & 0.9998 &	0.9998 & 0.9999 & 0.9999 & 0.9998 &	0.9998 \\
\hline
$\sigma$ Orionis &	0.9988 & 0.9988 & 0.9988 & 0.9988 &	0.9988 & 0.9988 & 0.9988 & 0.9988 &	0.9988 & 0.9988 \\
\hline
\end{tabular}
\end{table}

\begin{table}
\centering
\caption{ Same as Table \ref{tab:2} but for exponential LF and Z=0.019 }\label{tab:11}
    
\begin{tabular}{c|c|c|c|c|c|c|c|c|c|c}
\hline
Names & $\beta$ & $\beta$ & DE & DPL & E & E & G2 & G2 & G3 & G3 \\
of & (p, q) &  (p, q)  & ($T_{0}$, $\tau$) & ($\alpha$, $\beta$) & ($\lambda$) & ($\lambda$)  & ($\mu_{1}$, $\sigma_{1}$)  & ($\mu_{1}$, $\sigma_{1}$)  & ($\mu_{1}$, $\sigma_{1}$)  & ($\mu_{1}$, $\sigma_{1}$)  \\
observed & (2, 2) & (5, 1) & (15, 2) & ($\tau$) & (5) & (10) & ($\mu_{2}$, $\sigma_{2}$) & ($\mu_{2}$, $\sigma_{2}$) & ($\mu_{2}$, $\sigma_{2}$) & ($\mu_{2}$, $\sigma_{2}$)  \\
star  &  &  &  & (50, 10) &  &  & (q) & (q) & ($\mu_{3}$, $\sigma_{3}$) & ($\mu_{3}$, $\sigma_{3}$) \\
clusters  &  &  &  & (15) &  &  & (11, 0.2) & (3, 0.5) & ($q_{1}$, $q_{2}$) & ($q_{1}$, $q_{2}$) \\
 &  &  &  &  &  &  & (13, 0.5) & (8, 0.9)  & (2, 0.05) & (9, 0.05) \\
 &  &  &  &  &  &  & (0.5) & (0.6)  & (5, 0.05) & (11, 0.05) \\
 & &  &  &  & &  &   &  & (7, 0.05) & (13, 0.05) \\
 &  &  &  & &  &  &   &  & (0.5, 0.3) & (0.5, 0.3) \\
\hline
47 Tucanae &	0.8766 & 0.9506 & 0.9544 & 0.9572 &	0.9578 & 0.9586 & 0.9510 &	0.9487 & 0.8503 & 0.9541 \\
\hline
M 14 & 0.9640 &	0.9826 & 0.9802 & 0.9844 & 0.9908 &	0.9890 & 0.9800 & 0.9805 & 0.9654 &	0.9801 \\
\hline
M 5 & 0.8363 & 0.9336 &	0.9408 & 0.9433 & 0.9380 &	0.9379 & 0.9360 & 0.9405 & 0.8518 &	0.9415 \\
\hline
M 92 & 0.9207 &	0.9541 & 0.9752 & 0.9729 & 0.9560 &	0.9587 & 0.9609 & 0.9713 & 0.9184 &	0.9733 \\
\hline
M53 & 0.9634 & 0.9824 &	0.9838 & 0.9850 & 0.9805 & 0.9820 & 0.9810 & 0.9822 & 0.9686 &	0.9842 \\
\hline
NGC 1261 & 0.9912 &	0.9915 & 0.9963 & 0.9948 & 0.9916 &	0.9930 & 0.9952 & 0.9955 & 0.9948 &	0.9934 \\
\hline
NGC 1904 & 0.9323 & 0.9984 & 0.9485 & 0.9987 & 0.9985 & 0.9985 & 0.9982 & 0.9986 & 0.9348 & 0.9982 \\
\hline
NGC 2808 & 0.9599 &	0.9841 & 0.9811 & 0.9881 & 0.9859 &	0.9902 & 0.9866 & 0.9805 & 0.9642 &	0.9836 \\
\hline
NGC 288 & 0.9635 & 0.9852 &	0.9902 & 0.9914 & 0.9873 &	0.9908 & 0.9885 & 0.9906 & 0.9649 &	0.9876 \\
\hline
NGC 3201 & 0.9590 &	0.9872 & 0.9909 & 0.9919 & 0.9922 &	0.9933 & 0.9910 & 0.9859 & 0.9655 &	0.9873 \\
\hline
NGC 362 & 0.9668 & 0.9872 &	0.9910 & 0.9915 & 0.9937 &	0.9951 & 0.9905 & 0.9911 & 0.9669 &	0.9876 \\
\hline
NGC 4147 & 0.9631 &	0.9873 & 0.9960 & 0.9962 & 0.9938 &	0.9937 & 0.9932 & 0.9970 & 0.9644 &	0.9956 \\
\hline
NGC 4372 & 0.9644 &	0.9863 & 0.9803 & 0.9846 & 0.9901 &	0.9947 & 0.9860 & 0.9802 & 0.9654 &	0.9800 \\
\hline
NGC 4590 & 0.9575 &	0.9840 & 0.9886 & 0.9904 & 0.9867 &	0.9918 & 0.9873 & 0.9888 & 0.9655 &	0.9876 \\
\hline
NGC 4833 & 0.9645 &	0.9806 & 0.9858 & 0.9881 & 0.9881 &	0.9885 & 0.9832 & 0.9805 & 0.9655 &	0.9836 \\
\hline
NGC 5466 & 0.9578 &	0.9854 & 0.9922 & 0.9945 & 0.9851 &	0.9895 & 0.9910 & 0.9900 & 0.9657 &	0.9923 \\
\hline
NGC 5694 & 0.9924 &	0.9933 & 0.9978 & 0.9960 & 0.9973 &	0.9981 & 0.9968 & 0.9982 & 0.9822 &	0.9949 \\
\hline
NGC 5927 & 0.9649 &	0.9854 & 0.9800 & 0.9839 & 0.9933 &	0.9931 & 0.9838 & 0.9805 & 0.9654 &	0.9799 \\
\hline
NGC 6121 & 0.9647 &	0.9811 & 0.9749 & 0.9800 & 0.9869 &	0.9863 & 0.9772 & 0.9708 & 0.9654 &	0.9753 \\
\hline
NGC 6171 & 0.9650 &	0.9902 & 0.9877 & 0.9898 & 0.9972 &	0.9987 & 0.9890 & 0.9818 & 0.9634 &	0.9867 \\
\hline
NGC 6205 & 0.8608 &	0.9369 & 0.9558 & 0.9502 & 0.9383 &	0.9386 & 0.9411 & 0.9479 & 0.8702 &	0.9504 \\
\hline
NGC 6218 & 0.9582 &	0.9589 & 0.9585 & 0.9659 & 0.9640 &	0.9651 & 0.9610 & 0.9571 & 0.9435 &	0.9628 \\
\hline
NGC 6254 & 0.9600 &	0.9668 & 0.9641 & 0.9716 & 0.9721 &	0.9719 & 0.9669 & 0.9664 & 0.9655 &	0.9680 \\
\hline
NGC 6352 & 0.9670 &	0.9741 & 0.9676 & 0.9726 & 0.9810 &	0.9820 & 0.9729 & 0.9640 & 0.9592 &	0.9691 \\
\hline
NGC 6356 & 0.9680 &	0.9929 & 0.9812 & 0.9844 & 0.9990 &	0.9997 & 0.9900 & 0.9813 & 0.9660 &	0.9799 \\
\hline
NGC 6362 & 0.9915 & 0.9912 & 0.9942 & 0.9933 & 0.9951 &	0.9946 & 0.9938 & 0.9952 & 0.9670 &	0.9908 \\
\hline
NGC 6397 & 0.9587 &	0.9642 & 0.9645 & 0.9727 & 0.9702 &	0.9707 & 0.9646 & 0.9666 & 0.9655 &	0.9695 \\
\hline
NGC 6553 & 0.9628 &	0.9977 & 0.9872 & 0.9913 & 0.9975 &	0.9980 & 0.9965 & 0.9824 & 0.9657 &	0.9885 \\
\hline
NGC 6624 & 0.8551 & 0.9510 & 0.9700 & 0.9701 & 0.9558 &	0.9567 & 0.9586 & 0.9693 & 0.8490 &	0.9703 \\
\hline
NGC 6637 & 0.9630 &	0.9858 & 0.9858 & 0.9885 & 0.9930 &	0.9939 & 0.9866 & 0.9858 & 0.9655 &	0.9841 \\
\hline
NGC 6712 & 0.9669 &	0.9910 & 0.9879 & 0.9895 & 0.9974 &	0.9974 & 0.9881 & 0.9868 & 0.9654 &	0.9862 \\
\hline
NGC 6838 & 0.9095 &	0.9620 & 0.9536 & 0.9619 & 0.9617 &	0.9662 & 0.9518 & 0.9479 & 0.8539 &	0.9592 \\
\hline
NGC 7006 & 0.9912 &	0.9916 & 0.9944 & 0.9952 & 0.9945 &	0.9958 & 0.9937 & 0.9955 & 0.9719 &	0.9935 \\
\hline
NGC 7078 & 0.9204 &	0.9507 & 0.9744 & 0.9683 & 0.9529 &	0.9542 & 0.9608 & 0.9669 & 0.9275 &	0.9681 \\
\hline
NGC 7099 & 0.9613 &	0.9885 & 0.9958 & 0.9975 & 0.9912 &	0.9954 & 0.9948 & 0.9961 & 0.9659 &	0.9959 \\
\hline
NGC 7492 & 0.9911 &	0.9925 & 0.9955 & 0.9940 & 0.9960 &	0.9963 & 0.9958 & 0.9972 & 0.9928 &	0.9924 \\
\hline
NGC 2419 & 0.9582 &	0.9835 & 0.9844 & 0.9865 & 0.9865 & 0.9892 & 0.9850 & 0.9828 & 0.9640 &	0.9826 \\
\hline
NGC 5272 & 0.9420 &	0.9582 & 0.9722 & 0.9737 & 0.9618 &	0.9622 & 0.9614 & 0.9682 &	0.9193 & 0.9729 \\
\hline
$\omega$ Centauri & 0.9452 & 0.9652 & 0.9771 &	0.9777 &	0.9677 & 0.9687 & 0.9683 & 0.9736 &	0.9385 & 0.9735 \\
\hline
Palomar 4 &  0.9925 & 0.9914 & 0.9952 & 0.9940 & 0.9964 & 0.9979 & 0.9936 &	0.9909 & 0.9673 & 0.9923 \\
\hline

\end{tabular}
\end{table}

\begin{table}
\centering
\caption{ Same as Table \ref{tab:11} but for SMC observed star clusters }\label{tab:12}
    
\begin{tabular}{c|c|c|c|c|c|c|c|c|c|c}
\hline
Names & $\beta$ & $\beta$ & DE & DPL & E & E & G2 & G2 & G3 & G3 \\
of & (p, q) &  (p, q)  & ($T_{0}$, $\tau$) & ($\alpha$, $\beta$) & ($\lambda$) & ($\lambda$)  & ($\mu_{1}$, $\sigma_{1}$)  & ($\mu_{1}$, $\sigma_{1}$)  & ($\mu_{1}$, $\sigma_{1}$)  & ($\mu_{1}$, $\sigma_{1}$)  \\
observed & (2, 2) & (5, 1) & (15, 2) & ($\tau$) & (5) & (10) & ($\mu_{2}$, $\sigma_{2}$) & ($\mu_{2}$, $\sigma_{2}$) & ($\mu_{2}$, $\sigma_{2}$) & ($\mu_{2}$, $\sigma_{2}$)  \\
star  &  &  &  & (50, 10) &  &  & (q) & (q) & ($\mu_{3}$, $\sigma_{3}$) & ($\mu_{3}$, $\sigma_{3}$) \\
clusters  &  &  &  & (15) &  &  & (11, 0.2) & (3, 0.5) & ($q_{1}$, $q_{2}$) & ($q_{1}$, $q_{2}$) \\
 &  &  &  &  &  &  & (13, 0.5) & (8, 0.9)  & (2, 0.05) & (9, 0.05) \\
 &  &  &  &  &  &  & (0.5) & (0.6)  & (5, 0.05) & (11, 0.05) \\
 & &  &  &  & &  &   &  & (7, 0.05) & (13, 0.05) \\
 &  &  &  & &  &  &   &  & (0.5, 0.3) & (0.5, 0.3) \\
\hline
SMC Br\"{u}ck 2 & 0.9581 & 0.9749 &	0.9853 & 0.9885 &	0.9771 & 0.9783 & 0.9815 & 0.9850 & 0.9645 & 0.9858 \\
\hline
SMC Br\"{u}ck 4 & 0.9579 & 0.9714 &	0.9823 & 0.9834 &	0.9711 & 0.9754 & 0.9749 & 0.9784 &	0.9646 & 0.9815 \\
\hline
SMC Br\"{u}ck 6 & 0.9577 & 0.9695 &	0.9717 & 0.9774 &	0.9716 & 0.9725 & 0.9716 & 0.9709 &	0.9655 & 0.9739 \\
\hline
SMC HW 5 & 0.9590 &	0.9646 & 0.9727 & 0.9768 & 0.9707 &	0.9716 & 0.9687 & 0.9762 & 0.9647 &	0.9759 \\
\hline
SMC HW 6 & 0.9586 &	0.9710 & 0.9841 & 0.9845 & 0.9707 &	0.9753 & 0.9753 & 0.9795 & 0.9647 &	0.9825 \\
\hline
SMC Kron 11 & 0.9578 & 0.9649 &	0.9774 & 0.9799 &	0.9684 & 0.9687 & 0.9702 & 0.9812 &	0.9656 & 0.9806 \\
\hline
SMC Kron 8 & 0.9580 & 0.9789 & 0.9915 &	0.9911 &	0.9833 & 0.9819 & 0.9821 & 0.9903 &	0.9645 & 0.9894 \\
\hline
SMC Lindsay 14 & 0.9587 & 0.9671 & 0.9798 &	0.9804 &	0.9700 & 0.9724 & 0.9747 & 0.9798 &	0.9650 & 0.9795 \\
\hline
SMC NGC 152 & 0.9579 & 0.9785 &	0.9882 & 0.9898 &	0.9780 & 0.9818 & 0.9807 & 0.9847 &	0.9655 & 0.9874 \\
\hline

\end{tabular}
\end{table}

\begin{table}
\centering
\caption{ Same as Table \ref{tab:11} but for MW observed open clusters}\label{tab:13}
    
\begin{tabular}{c|c|c|c|c|c|c|c|c|c|c}
\hline
Names & $\beta$ & $\beta$ & DE & DPL & E & E & G2 & G2 & G3 & G3 \\
of & (p, q) &  (p, q)  & ($T_{0}$, $\tau$) & ($\alpha$, $\beta$) & ($\lambda$) & ($\lambda$)  & ($\mu_{1}$, $\sigma_{1}$)  & ($\mu_{1}$, $\sigma_{1}$)  & ($\mu_{1}$, $\sigma_{1}$)  & ($\mu_{1}$, $\sigma_{1}$)  \\
observed & (2, 2) & (5, 1) & (15, 2) & ($\tau$) & (5) & (10) & ($\mu_{2}$, $\sigma_{2}$) & ($\mu_{2}$, $\sigma_{2}$) & ($\mu_{2}$, $\sigma_{2}$) & ($\mu_{2}$, $\sigma_{2}$)  \\
star  &  &  &  & (50, 10) &  &  & (q) & (q) & ($\mu_{3}$, $\sigma_{3}$) & ($\mu_{3}$, $\sigma_{3}$) \\
clusters  &  &  &  & (15) &  &  & (11, 0.2) & (3, 0.5) & ($q_{1}$, $q_{2}$) & ($q_{1}$, $q_{2}$) \\
 &  &  &  &  &  &  & (13, 0.5) & (8, 0.9)  & (2, 0.05) & (9, 0.05) \\
 &  &  &  &  &  &  & (0.5) & (0.6)  & (5, 0.05) & (11, 0.05) \\
 & &  &  &  & &  &   &  & (7, 0.05) & (13, 0.05) \\
 &  &  &  & &  &  &   &  & (0.5, 0.3) & (0.5, 0.3) \\
\hline
H and $\chi$ Persei & 0.5730 & 0.8124 &	0.8124 & 0.8083 &	0.8147 & 0.8147 & 0.8078 & 0.8074 &	0.5613 & 0.8128 \\
\hline
NGC 2264 & 0.3163 &	0.6651 & 0.6684 & 0.6618 & 0.6685 &	0.6653 & 0.6620 & 0.6675 & 0.3136 &	0.6668 \\
\hline
NGC 2547 & 0.4264 &	0.6659 & 0.6863 & 0.6735 & 0.6575 &	0.6652 & 0.6636 & 0.6664 & 0.4566 &	0.6830 \\
\hline
NGC 6811 & 0.9642 &	0.9828 & 0.9869 & 0.9891 & 0.9822 &	0.9883 & 0.9805 & 0.9760 & 0.9654 &	0.9849 \\
\hline
NGC 7160 & 0.9998 &	0.9999 & 1.0000 & 0.9999 & 1.0000 &	0.9999 & 0.9999 & 0.9999 & 0.9998 &	0.9999 \\
\hline
$\sigma$ Orionis &	0.9988 & 0.9988 & 0.9988 & 0.9988 &	0.9988 & 0.9988 & 0.9988 & 0.9988 &	0.9988 & 0.9988 \\
\hline
\end{tabular}
\end{table}

\begin{table}
\centering
\caption{ Same as Table \ref{tab:11} for TPL ($\alpha$ = 1.05, $\beta$ = 8, $\nu$ = 2) LF and Z=0.019}\label{tab:14}
    
\begin{tabular}{c|c|c|c|c|c|c|c|c|c|c}
\hline
Names & $\beta$ & $\beta$ & DE & DPL & E & E & G2 & G2 & G3 & G3 \\
of & (p, q) &  (p, q)  & ($T_{0}$, $\tau$) & ($\alpha$, $\beta$) & ($\lambda$) & ($\lambda$)  & ($\mu_{1}$, $\sigma_{1}$)  & ($\mu_{1}$, $\sigma_{1}$)  & ($\mu_{1}$, $\sigma_{1}$)  & ($\mu_{1}$, $\sigma_{1}$)  \\
observed & (2, 2) & (5, 1) & (15, 2) & ($\tau$) & (5) & (10) & ($\mu_{2}$, $\sigma_{2}$) & ($\mu_{2}$, $\sigma_{2}$) & ($\mu_{2}$, $\sigma_{2}$) & ($\mu_{2}$, $\sigma_{2}$)  \\
star  &  &  &  & (50, 10) &  &  & (q) & (q) & ($\mu_{3}$, $\sigma_{3}$) & ($\mu_{3}$, $\sigma_{3}$) \\
clusters  &  &  &  & (15) &  &  & (11, 0.2) & (3, 0.5) & ($q_{1}$, $q_{2}$) & ($q_{1}$, $q_{2}$) \\
 &  &  &  &  &  &  & (13, 0.5) & (8, 0.9)  & (2, 0.05) & (9, 0.05) \\
 &  &  &  &  &  &  & (0.5) & (0.6)  & (5, 0.05) & (11, 0.05) \\
 & &  &  &  & &  &   &  & (7, 0.05) & (13, 0.05) \\
 &  &  &  & &  &  &   &  & (0.5, 0.3) & (0.5, 0.3) \\
\hline
47 Tucanae &	0.6556 & 0.6078 & 0.5795 & 0.5882 &	0.6971 & 0.7119 & 0.6392 &	0.5618 & 0.5697 & 0.6229 \\
\hline
M 14 & 0.8550 &	0.8227 & 0.7676 & 0.7743 & 0.8938 &	0.8929 & 0.7938 & 0.7728 & 0.7674 &	0.7827 \\
\hline
M 5 & 0.5712 & 0.5344 &	0.6001 & 0.5960 & 0.6290 &	0.6474 & 0.6821 & 0.5769 & 0.5745 &	0.6688 \\
\hline
M 92 & 0.7090 &	0.7307 & 0.8459 & 0.8346 & 0.7372 &	0.7547 & 0.9084 & 0.8348 & 0.8317 &	0.9034 \\
\hline
M 53 & 0.8704 &	0.8695 & 0.8640 & 0.8598 & 0.8770 &	0.8761 & 0.9029 & 0.8670 & 0.8585 &	0.8915 \\
\hline
NGC 1261 & 0.8184 &	0.8169 & 0.8598 & 0.8555 & 0.8679 &	0.8860 & 0.8854 & 0.8595 & 0.8620 &	0.8820 \\
\hline
NGC 1904 & 0.9973 &	0.9973 & 0.9973 & 0.9973 & 0.9973 &	0.9973 & 0.9973 & 0.9973 & 0.9973 &	0.9973 \\
\hline
NGC 2808 & 0.8508 &	0.7930 & 0.7230 & 0.7385 & 0.9031 &	0.9122 & 0.7420 & 0.7248 & 0.7288 &	0.7344 \\
\hline
NGC 288 & 0.8238 & 0.7920 &	0.8011 & 0.8076 & 0.8755 &	0.8944 & 0.8327 & 0.7935 & 0.7995 &	0.8257 \\
\hline
NGC 3201 & 0.8713 &	0.8228 & 0.8041 & 0.8095 & 0.9164 &	0.9301 & 0.8182 & 0.7990 & 0.8002 &	0.8110 \\
\hline
NGC 362 & 0.8858 & 0.8232 &	0.8074 & 0.8143 & 0.9452 &	0.9604 & 0.8306 & 0.7948 & 0.7969 &	0.8232 \\
\hline
NGC 4147 & 0.8332 &	0.8071 & 0.9074 & 0.9038 & 0.8646 &	0.8831 & 0.9392 & 0.8913 & 0.8944 &	0.9351 \\
\hline
NGC 4372 & 0.9053 &	0.8454 & 0.7378 & 0.7569 & 0.9483 &	0.9576 & 0.7389 & 0.7468 & 0.7494 &	0.7283 \\
\hline
NGC 4590 & 0.7789 &	0.7446 & 0.8418 & 0.8382 & 0.8463 &	0.8569 & 0.8834 & 0.8244 & 0.8243 &	0.8778 \\
\hline
NGC 4833 & 0.8399 &	0.7824 & 0.7217 & 0.7340 & 0.8946 &	0.9002 & 0.7512 & 0.7194 & 0.7195 &	0.7389 \\
\hline
NGC 5466 & 0.8062 &	0.7830 & 0.9098 & 0.9006 & 0.8516 &	0.8635 & 0.9550 & 0.8864 & 0.8882 &	0.9530 \\
\hline
NGC 5694 & 0.8918 &	0.8756 & 0.9216 & 0.9233 & 0.9399 &	0.9516 & 0.9315 & 0.9141 & 0.9172 &	0.9307 \\
\hline
NGC 5927 & 0.9433 &	0.9222 & 0.8020 & 0.8239 & 0.9639 &	0.9697 & 0.7872 & 0.8250 & 0.8349 &	0.7756 \\
\hline
NGC 6121 & 0.9135 &	0.8818 & 0.7482 & 0.7736 & 0.9364 &	0.9412 & 0.7168 & 0.7739 & 0.7828 &	0.7128 \\
\hline
NGC 6171 & 0.9575 &	0.9202 & 0.8215 & 0.8361 & 0.9822 &	0.9871 & 0.8128 & 0.8300 & 0.8310 &	0.8063 \\
\hline
NGC 6205 & 0.6209 &  0.6216 & 0.7052 & 0.6960 &	0.6673 & 0.6862 & 0.7731 &	0.6966 & 0.7003 & 0.7643 \\
\hline
NGC 6218 & 0.6629 &	0.6580 & 0.6171 & 0.6242 & 0.7021  & 0.7077 & 0.6634 &	0.6224 & 0.6195 & 0.6535 \\
\hline
NGC 6254 & 0.7657 &	0.7348 & 0.6350 & 0.6574 & 0.8037 &	0.8098 & 0.6420 & 0.6540 & 0.6642 &	0.6331 \\
\hline
NGC 6352 & 0.9078 &	0.8738 & 0.7459 & 0.7722 & 0.9351 &	0.9418 & 0.7165 & 0.7671 & 0.7750 &	0.7160 \\
\hline
NGC 6356 & 0.9538 &	0.9157 & 0.7813 & 0.7993 & 0.9838 &	0.9935 & 0.7617 & 0.7977 & 0.8042 &	0.7569 \\
\hline
NGC 6362 & 0.8838 &	0.8543 & 0.8372 & 0.8415 & 0.9247 &	0.9412 & 0.8574 & 0.8323 & 0.8359 &	0.8509 \\
\hline
NGC 6397 & 0.6520 &	0.6386 & 0.6137 & 0.6190 & 0.7013 &	0.7109 & 0.6697 & 0.6191 & 0.6144 &	0.6538 \\
\hline
NGC 6553 & 0.9806 &	0.9792 & 0.9394 & 0.9485 & 0.9873 &	0.9889 & 0.9205 & 0.9492 & 0.9578 &	0.9265 \\
\hline
NGC 6624 & 0.6366 &	0.6304 & 0.7295 & 0.7198 & 0.6606 &	0.6681 & 0.8198 & 0.7037 & 0.6992 &	0.8051 \\
\hline
NGC 6637 & 0.9263 &	0.8811 & 0.7881 & 0.8032 & 0.9576 &	0.9671 & 0.7897 & 0.7935 & 0.7999 &	0.7861 \\
\hline
NGC 6712 & 0.9428 &	0.9102 & 0.8135 & 0.8264 & 0.9721 &	0.9763 & 0.8094 & 0.8246 & 0.8189 &	0.8002 \\
\hline
NGC 6838 & 0.7654 &	0.7468 & 0.6284 & 0.6477 & 0.7801 &	0.7867 & 0.6221 & 0.6386 & 0.6562 &	0.6247 \\
\hline
NGC 7006 & 0.8574 &	0.8286 & 0.8748 & 0.8739 & 0.9001 &	0.9168 & 0.8978 & 0.8650 & 0.8682 &	0.8928 \\
\hline
NGC 7078 & 0.6883 &	0.6836 & 0.8087 & 0.7979 & 0.7337 &	0.7524 & 0.8779 & 0.7941 & 0.7903 &	0.8688 \\
\hline
NGC 7099 & 0.8463 &	0.8246 & 0.9386 & 0.9380 & 0.8824 &	0.8985 & 0.9648 & 0.9223 & 0.9224 &	0.9633 \\
\hline
NGC 7492 & 0.8495 &	0.8440 & 0.8891 & 0.8877 & 0.8887 &	0.9025 & 0.9088 & 0.8822 & 0.8834 &	0.9064 \\
\hline
NGC 2419 & 0.7895 &	0.7551 & 0.7988 & 0.8007 & 0.8432 &	0.8614 & 0.8335 & 0.7870 & 0.7906 &	0.8282 \\
\hline
NGC 5272 & 0.6652 &	0.6517 & 0.7479 & 0.7367 & 0.7199 &	0.7387 & 0.8191 & 0.7296 & 0.7290 &	0.8116 \\
\hline
$\omega$ Centauri & 0.7327 & 0.7183 & 0.8131 &	0.8067 &	0.7726 & 0.7848 & 0.8721 & 0.7973 &	0.8010 & 0.8690 \\
\hline
Palomar 4 &	0.8768 & 0.8185 & 0.8471 & 0.8514 &	0.9465 & 0.9616 & 0.8557 &	0.8334 & 0.8370 & 0.8527 \\
\hline

\end{tabular}
\end{table}

\begin{table}
\centering
\caption{Same as Table \ref{tab:14} but for SMC observed star clusters }\label{tab:15}
    
\begin{tabular}{c|c|c|c|c|c|c|c|c|c|c}
\hline
Names & $\beta$ & $\beta$ & DE & DPL & E & E & G2 & G2 & G3 & G3 \\
of & (p, q) &  (p, q)  & ($T_{0}$, $\tau$) & ($\alpha$, $\beta$) & ($\lambda$) & ($\lambda$)  & ($\mu_{1}$, $\sigma_{1}$)  & ($\mu_{1}$, $\sigma_{1}$)  & ($\mu_{1}$, $\sigma_{1}$)  & ($\mu_{1}$, $\sigma_{1}$)  \\
observed & (2, 2) & (5, 1) & (15, 2) & ($\tau$) & (5) & (10) & ($\mu_{2}$, $\sigma_{2}$) & ($\mu_{2}$, $\sigma_{2}$) & ($\mu_{2}$, $\sigma_{2}$) & ($\mu_{2}$, $\sigma_{2}$)  \\
star  &  &  &  & (50, 10) &  &  & (q) & (q) & ($\mu_{3}$, $\sigma_{3}$) & ($\mu_{3}$, $\sigma_{3}$) \\
clusters  &  &  &  & (15) &  &  & (11, 0.2) & (3, 0.5) & ($q_{1}$, $q_{2}$) & ($q_{1}$, $q_{2}$) \\
 &  &  &  &  &  &  & (13, 0.5) & (8, 0.9)  & (2, 0.05) & (9, 0.05) \\
 &  &  &  &  &  &  & (0.5) & (0.6)  & (5, 0.05) & (11, 0.05) \\
 & &  &  &  & &  &   &  & (7, 0.05) & (13, 0.05) \\
 &  &  &  & &  &  &   &  & (0.5, 0.3) & (0.5, 0.3) \\
\hline
SMC Br\"{u}ck 2 & 0.7555 & 0.7492 &	0.8649 & 0.8456 &	0.7916 & 0.7886 & 0.9401 & 0.8475 &	0.8378 & 0.9383 \\
\hline
SMC Br\"{u}ck 4 & 0.7204 & 0.7092 &	0.8291 & 0.8078 &	0.7537 & 0.7519 & 0.9188 & 0.8091 &	0.7931 & 0.9052 \\
\hline
SMC Br\"{u}ck 6 & 0.6944 & 0.6809 &	0.7308 & 0.7250 &	0.7408 & 0.7415 & 0.7855 & 0.7329 &	0.7292 & 0.7786 \\
\hline
SMC HW 5 & 0.6857 &	0.6650 & 0.7713 & 0.7566 & 0.7296 &	0.7341 & 0.8614 & 0.7486 & 0.7359 &	0.8469 \\
\hline
SMC HW 6 & 0.7155 &	0.6964 & 0.8298 & 0.8103 & 0.7589 &	0.7583 & 0.9201 & 0.8076 & 0.7918 & 0.9066 \\ 
\hline
SMC Kron 11 & 0.6574 & 0.6616 &	0.7920 & 0.7732 &	0.6898 & 0.6912 & 0.8893 & 0.7736 &	0.7614 & 0.8757 \\
\hline
SMC Kron 8 & 0.7618 & 0.7411 & 0.8798 &	0.8632 &	0.8059 & 0.8080 & 0.9548 & 0.8598 &	0.8559 & 0.9534 \\
\hline
SMC Lindsay 14 & 0.7047 & 0.6766 & 0.8051 &	0.7842 &	0.7565 & 0.7589 & 0.8984 & 0.7798 &	0.7633 & 0.8835 \\
\hline
SMC NGC 152 & 0.7278 & 0.7093 &	0.8507 & 0.8320 &	0.7795 & 0.7800 & 0.9298 & 0.8297 &	0.8213 & 0.9275 \\
\hline
\end{tabular}
\end{table}

\begin{table}
\centering
\caption{ Same as Table \ref{tab:14} but for MW observed open clusters}\label{tab:16}
    
\begin{tabular}{c|c|c|c|c|c|c|c|c|c|c}
\hline
Names & $\beta$ & $\beta$ & DE & DPL & E & E & G2 & G2 & G3 & G3 \\
of & (p, q) &  (p, q)  & ($T_{0}$, $\tau$) & ($\alpha$, $\beta$) & ($\lambda$) & ($\lambda$)  & ($\mu_{1}$, $\sigma_{1}$)  & ($\mu_{1}$, $\sigma_{1}$)  & ($\mu_{1}$, $\sigma_{1}$)  & ($\mu_{1}$, $\sigma_{1}$)  \\
observed & (2, 2) & (5, 1) & (15, 2) & ($\tau$) & (5) & (10) & ($\mu_{2}$, $\sigma_{2}$) & ($\mu_{2}$, $\sigma_{2}$) & ($\mu_{2}$, $\sigma_{2}$) & ($\mu_{2}$, $\sigma_{2}$)  \\
star  &  &  &  & (50, 10) &  &  & (q) & (q) & ($\mu_{3}$, $\sigma_{3}$) & ($\mu_{3}$, $\sigma_{3}$) \\
clusters  &  &  &  & (15) &  &  & (11, 0.2) & (3, 0.5) & ($q_{1}$, $q_{2}$) & ($q_{1}$, $q_{2}$) \\
 &  &  &  &  &  &  & (13, 0.5) & (8, 0.9)  & (2, 0.05) & (9, 0.05) \\
 &  &  &  &  &  &  & (0.5) & (0.6)  & (5, 0.05) & (11, 0.05) \\
 & &  &  &  & &  &   &  & (7, 0.05) & (13, 0.05) \\
 &  &  &  & &  &  &   &  & (0.5, 0.3) & (0.5, 0.3) \\
\hline
H and $\chi$ Persei & 0.5822 & 0.5748 &	0.5741	& 0.5760 &	0.5919 & 0.5910 & 0.5854 & 0.5705 &	0.5747 & 0.5857 \\
\hline
NGC 2264 & 0.7792 &	0.7783 & 0.7731 & 0.7744 & 0.7783 &	0.7792 & 0.7780 & 0.7750 & 0.7761 &	0.7791 \\
\hline
NGC 2547 & 0.5293 &	0.5394 & 0.5798 & 0.5634 & 0.5101 &	0.5087 & 0.6417 & 0.5798 & 0.5741 &	0.6365 \\
\hline
NGC 6811 & 0.8134 &	0.7511 & 0.8036 & 0.7988 & 0.8657 &	0.8661 & 0.8494 & 0.7808 & 0.7788 &	0.8479 \\
\hline
NGC 7160 & 0.9999 &	0.9999 & 0.9999 & 0.9999 & 0.9999 &	0.9999 & 0.9999 & 0.9999 & 0.9999 &	0.9999 \\
\hline
$\sigma$ Orionis &	0.9988 & 0.9988 & 0.9988 & 0.9988 &	0.9988 & 0.9988 & 0.9988 & 0.9988 &	0.9988 & 0.9988 \\
\hline

\end{tabular}
\end{table}

\acknowledgments
One of the authors S.M. is very much grateful to UGC,  India, for approving a  JRF grant for the work.
Author S.P. acknowledges INSPIRE JRF Vide sanction Order No. DST/INSPIRE Fellowship/2017/IF170368 under the Department of Science and Technology (DST) INSPIRE program, Government of India.

\clearpage

\bibliographystyle{aasjournal}
\bibliography{MS}

\appendix 

\begin{table}[h]
\centering
\renewcommand{\arraystretch}{0.95}
\caption{Metallicities and Galacto-centric distances of various globular clusters\footnote{\url{http://www.naic.edu/~pulsar/catalogs/mwgc.txt}} in MW.}\label{tab:17}
    
\begin{tabular}{  c  |  c  |  c  }
\hline
Names of observed  &  Metallicity  & Galacto-centric \\
globular clusters &  [Fe/H] & distances (kpc) \\
\hline
47 Tucanae & $-$ 0.76 & 7.4  \\
\hline
M 14 & $-$ 1.39 &  4.1 \\
\hline
M 5 & $-$ 1.27 &  6.2 \\
\hline
M 92 & $-$ 2.28 &  9.6 \\
\hline
M 53 & $-$ 1.99 &  18.3 \\
\hline
NGC 1261  & $-$ 1.35 & 18.2  \\
\hline
NGC 1904 & $-$ 1.57 &  18.8 \\
\hline
NGC 2808 & $-$ 1.15 & 11.1  \\
\hline
NGC 288 & $-$ 1.24 &  12.0 \\
\hline
NGC 3201 & $-$ 1.58 &  8.9 \\
\hline
NGC 362 & $-$ 1.16 &  9.4 \\
\hline
NGC 4147 & $-$ 1.83 &  21.3 \\
\hline
NGC 4372 & $-$ 2.09 &  7.1  \\
\hline
NGC 4590 & $-$ 2.06 &  10.1 \\
\hline
NGC 4833 & $-$ 1.80 &  7.0 \\
\hline
NGC 5466 & $-$ 2.22 &  16.2 \\
\hline
NGC 5694 & $-$ 1.86 & 29.1  \\
\hline
NGC 5927 & $-$ 0.37 &  4.5 \\
\hline
NGC 6121 & $-$ 1.20 &  5.9 \\
\hline
NGC 6171 & $-$ 1.04 &  3.3 \\
\hline
NGC 6205 & $-$ 1.54 &  8.7 \\
\hline
NGC 6218 & $-$ 1.48 &  4.5 \\
\hline
NGC 6254 & $-$ 1.52 &  4.6 \\
\hline
NGC 6352 & $-$ 0.70 & 3.3  \\
\hline
NGC 6356 & $-$ 0.50 &  7.6 \\
\hline
NGC 6362 & $-$ 0.95 &  5.1 \\
\hline
NGC 6397 & $-$ 1.95 &  6.0 \\
\hline
NGC 6553 & $-$ 0.21 &  2.2 \\
\hline
NGC 6624 & $-$ 0.44 &  1.2 \\
\hline
NGC 6637 & $-$ 0.70 &  1.9 \\
\hline
NGC 6712 & $-$ 1.01 & 3.5  \\
\hline
NGC 6838 & $-$ 0.73 &   6.7 \\
\hline
NGC 7006 & $-$ 1.63 & 38.8  \\
\hline
NGC 7078 & $-$ 2.26 & 10.4  \\
\hline
NGC 7099 & $-$ 2.12 & 7.1  \\
\hline
NGC 7492 & $-$ 1.51 &  24.9 \\
\hline
NGC 2419 & $-$ 2.12 & 91.5  \\
\hline
NGC 5272 & $-$ 1.57 &  12.2 \\
\hline
$\omega$ Centauri & $-$ 1.62 &  6.4 \\
\hline
Palomar 4 & $-$ 1.48 & 111.8  \\
\hline
\end{tabular}
\end{table}

\end{document}